\documentclass[envcountsect]{llncs}

\usepackage[sumlimits]{amsmath}
\usepackage{emlines}
\usepackage{emlines2}

\begin{document}

\title{On the class of languages recognizable by 1-way quantum
finite automata}

\author{Andris Ambainis\inst{1}, Arnolds \c Kikusts\inst{2}, 
M\=aris Valdats\inst{2}}

\institute{
 Computer Science Division,
 University of California, Berkeley, CA94720, USA,
 \email{ambainis@cs.berkeley.edu}
\thanks{%
Research supported by Berkeley Fellowship for Graduate Studies and, in part,
NSF Grant CCR-9800024.}\and
 Institute of Mathematics and Computer Science,
 University of Latvia, Rai\c na bulv. 29, R\=\i ga,
Latvia\thanks{%
Research supported by Grant No.96.0282 from the
Latvian Council of Science and European Commission,
contract IST-1999-11234.}\\
 \email{sd70053@lanet.lv, sd70066@lanet.lv}
}

\def \lket {\left|}
\def \rket {\right\rangle}
\newcommand{\ket}[1]{\lket #1\rket}
\newcommand{\comment}[1]{}
\def \cin {\!\!\in\!\!}
\def \cnotin {\!\!\notin\!\!}
\def \Qspace {l_2(Q)}

\maketitle

\begin{abstract}
It is an open problem to characterize the class of languages
recognized by quantum finite automata (QFA).
We examine some necessary and some sufficient
conditions for a (regular) language to be recognizable
by a QFA. For a subclass of regular languages we get
a condition which is necessary and sufficient.

Also, we prove that the class of languages recognizable
by a QFA is not closed under union or any
other binary Boolean operation where both arguments
are significant.
\end{abstract}

\section{Introduction}

A quantum finite automaton (QFA) is a theoretical model for a quantum
computer with a finite memory.

If we compare them with their classical (non-quantum) counterparts,
QFAs have both strengths and weaknesses.
The strength of QFAs is shown by the fact that quantum automata can be exponentially more
space efficient than deterministic or probabilistic automata \cite{AF 98}.
The weakness of QFAs is caused by the fact that any quantum process has to be 
reversible (unitary).
This makes quantum automata unable to recognize some regular languages.

The first result of this type was obtained by Kondacs and Watrous \cite{KW 97} who 
showed that there is a language that can be recognized 
by a deterministic finite automaton (DFA) but cannot be recognized by QFA.
Later, Brodsky and Pippenger \cite{BP 99} generalized the construction of \cite{KW 97}
and showed that any regular language that does not satisfy the partial order condition
cannot be recognized by a QFA.
They also conjectured that all regular languages satisfying the partial order condition
can be recognized by a QFA.

In this paper, we disprove their conjecture. 
We show that, for a language to be recognizable by a 1-way QFA, its minimal
deterministic automaton must not contain several ``forbidden fragments".
One of these fragments is equivalent to the automaton not satisfying the partial
order condition. The other fragments are new.

A somewhat surprising feature of our ``forbidden fragments"
is that they consist of several parts (corresponding to different
beginnings of the word) and the language corresponding to every one
of them can be recognized but one cannot simultaneously recognize the
whole language without violating unitarity.

Our result implies that the set of languages recognizable by QFAs is
not closed under union. In particular, if we consider the language 
consisting of all words in the alphabet $\{a, b\}$ that have an even number 
of $a$'s after the first $b$, this language is not recognizable by a QFA,
although it is a union of two recognizable languages. (The first language
consists of all words with an even number of $a$'s before the first $b$ and 
an even number of $a$'s after the first $b$, the second language consists
of all words with an odd number of $a$'s before the first $b$ and
an even number of $a$'s after it.)
This answers a question of Brodsky and Pippenger \cite{BP 99}.

For a subclass of regular languages (languages that do not contain 
"two cycles in a row" construction shown in Fig. \ref{Bilde10}), 
we show that our conditions are necessary
and sufficient for a language to be recognizable by a QFA.
For arbitrary regular languages, we only know that these conditions
are necessary but we do not know if all languages satisfying them can
be recognized by a QFA.

\begin{subsection}{Definitions}

Quantum finite automata (QFA) were introduced independently by 
Moore and Crutchfield \cite{CM 97} 
and Kondacs and Watrous \cite{KW 97}. In this paper,
we consider the more general definition of QFAs \cite{KW 97}
(which includes the definition of \cite{CM 97} as a special case).

\begin{definition}
\label{def1}
A QFA is a tuple
$M=(Q;\Sigma ;V ;q_{0};Q_{acc};Q_{rej})$ where $Q$ is a finite set
of states, $\Sigma $ is an input alphabet, $V$ is a transition function,
$q_{0}\cin Q$ is a starting state, and $Q_{acc}\subseteq Q$
and $Q_{rej}\subseteq Q$
are sets of accepting and rejecting states
($Q_{acc}\cap Q_{rej}=\emptyset$).
The states in $Q_{acc}$ and $Q_{rej}$,
are called {\em halting states} and
the states in $Q_{non}=Q-(Q_{acc}\cup Q_{rej})$ are called
{\em non halting states}.
$\kappa$ and $\$$ are symbols that do not belong to $\Sigma$.
We use $\kappa$ and $\$$ as the left and the right endmarker,
respectively. The {\em working alphabet} of
$M$ is $\Gamma = \Sigma \cup \{\kappa ;\$\}$.

The state of $M$ can be any superposition of states in $Q$
(i. e., any linear combination of them with complex coefficients). 
We use $\ket{q}$ to denote the superposition consisting
of state $q$ only.
$l_2(Q)$ denotes the linear space consisting of all superpositions, with
$l_2$-distance on this linear space. 

The transition function $V$ is a mapping from $\Gamma\times \Qspace$
to $\Qspace$ such that, for every $a\cin\Gamma$, the function
$V_a:\Qspace\rightarrow\Qspace$ defined by $V_a(x)=V(a, x)$ is a 
unitary transformation (a linear transformation on $l_2(Q)$ that
preserves $l_2$ norm).
\end{definition}

The computation of a QFA starts in the superposition $|q_{0}\rangle$.
Then transformations corresponding to the left endmarker $\kappa$,
the letters of the input word $x$ and the right endmarker $\$$ are
applied. The transformation corresponding to $a\cin \Gamma$ consists
of two steps.

1. First, $V_{a}$ is applied. The new superposition $\psi^{\prime}$
is $V_{a}(\psi)$ where $\psi$ is the superposition before this step.

2. Then, $\psi^{\prime}$ is observed with respect to 
$E_{acc}, E_{rej}, E_{non}$ where
$E_{acc}=span\{|q\rangle:q\cin Q_{acc}\}$,
$E_{rej}=span\{|q\rangle :q\cin Q_{rej}\}$,
$E_{non}=span\{|q\rangle :q\cin Q_{non}\}$.
It means that if the system's state before the measurement was
$$\psi' = \sum_{q_i\in Q_{acc}} \alpha_i \ket{q_i} +
\sum_{q_j\in Q_{rej}} \beta_j \ket{q_j} +
\sum_{q_k\in Q_{non}} \gamma_k \ket{q_k}$$
then the measurement accepts $\psi'$ with probability $\Sigma\alpha_i^2$,
rejects with probability $\Sigma\beta_j^2$ and
continues the computation (applies transformations 
corresponding to next letters) with probability $\Sigma\gamma_k^2$
with the system having state $\psi=\Sigma\gamma_k\ket{q_k}$.

We regard these two transformations as reading a letter $a$.
We use $V'_a$ to denote the transformation consisting of
$V_a$ followed by projection to $E_{non}$.
This is the transformation mapping $\psi$ to the non-halting part
of $V_a(\psi)$. 
We use $V_w'$ to denote the product of transformations
$V_w'=V_{a_n}'V_{a_{n-1}}'\dots V_{a_2}'V_{a_1}'$,
where $a_i$ is the $i$-th letter of the word $w$.
We also use $\psi_y$ to denote the non-halting part of 
QFA's state after reading the left endmarker $\kappa$ and the
word $y\cin\Sigma^*$.
From the notation it follows that $\psi_w=V_{\kappa w}'(\ket{q_0})$.

We will say that an automaton recognizes a language $L$ with probability $p$
$(p>\frac{1}{2})$ if it accepts any word $x\cin L$ with
probability $\geq p$ and
rejects any word $x\cnotin L$ with probability $\geq p$.

\end{subsection}

\begin{subsection}{Previous work}

The previous work on 1-way quantum finite automata (QFAs) has mainly
considered 3 questions:
\begin{enumerate}
\item
What is the class of languages recognized by QFAs?
\item
What accepting probabilities can be achieved?
\item
How does the size of QFAs (the number of states) compare
to the size of deterministic (probabilistic) automata?
\end{enumerate}

In this paper, we consider the first question.
The first results in this direction were obtained by
Kondacs and Watrous \cite{KW 97}.

\begin{theorem}
\cite{KW 97}
\label{KWTheorem}
\begin{enumerate}
\item
All languages recognized by 1-way QFAs are regular.
\item
There is a regular language that cannot be recognized
by a 1-way QFA with probability $\frac{1}{2}+\epsilon$
for any $\epsilon>0$.
\end{enumerate}
\end{theorem}

Brodsky and Pippenger \cite{BP 99} generalized the second part
of Theorem \ref{KWTheorem} by showing that any language satisfying
a certain property is not recognizable by a QFA.

\begin{theorem}
\label{T13}
\cite{BP 99}
Let $L$ be a language and $M$ be its minimal automaton
(the smallest DFA recognizing $L$).
Assume that there is a word $x$ such that $M$  
contains states $q_1$, $q_2$ satisfying:
\begin{enumerate}
\item
$q_1\neq q_2$,
\item
If $M$ starts in the state $q_1$ and reads $x$,
it passes to $q_2$,
\item
If $M$ starts in the state $q_2$ and reads $x$,
it passes to $q_2$, and
\item
There is a word $y$ such that if M starts in $q_2$ and reads y, it passes to $q_1$,
\end{enumerate}
then $L$ cannot be recognized by any 1-way quantum finite automaton (Fig.\ref{Bilde1a}).
\end{theorem}

\hspace{0.25\textwidth}
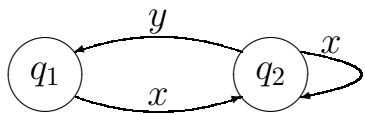
\begin{figure}[htb]
  \centering
  \begin{Large}
\unitlength 1.00mm
\linethickness{0.4pt}
\begin{picture}(70.00,17.00)
\put(20.00,10.00){\circle{10.00}}
\put(50.00,10.00){\circle{10.00}}
\put(20.00,10.00){\makebox(0,0)[cc]{$q_1$}}
\put(50.00,10.00){\makebox(0,0)[cc]{$q_2$}}
\put(54.00,7.00){\vector(-4,-1){0.2}}
\bezier{132}(54.00,13.00)(70.00,10.00)(54.00,7.00)
\put(46.00,7.00){\vector(3,1){0.2}}
\bezier{92}(24.00,7.00)(35.00,3.00)(46.00,7.00)
\put(35.00,7.00){\makebox(0,0)[cc]{$x$}}
\put(35.00,17.00){\makebox(0,0)[cc]{$y$}}
\put(58.00,14.00){\makebox(0,0)[cc]{$x$}}
\put(24.00,13.00){\vector(-3,-1){0.2}}
\bezier{92}(46.00,13.00)(35.00,17.00)(24.00,13.00)
\end{picture}

  \end{Large}
  \caption{Conditions of theorem \ref{T13}}
  \label{Bilde1a}
\end{figure}
\hspace{0.25\textwidth}

A language $L$ with the minimal automaton not containing
a fragment of Theorem \ref{T13} is called {\em satisfying
the partial order condition} \cite{MT 69}.
\cite{BP 99} conjectured that any language satisfying the partial
order condition is recognizable by a 1-way QFA.
In this paper, we disprove this conjecture.

Another direction of research is studying the accepting
probabilities of QFAs.
This direction started with Ambainis and Freivalds \cite{AF 98}
showing that the language $a^{*}b^{*}$ is recognizable by
a QFA with probability 0.68... but not with probability
$7/9+\epsilon$ for any $\epsilon>0$.
This showed that the classes of languages recognizable with
different probabilities are different.
Next results in this direction were obtained by \cite{ABFK 99}
who studied the probabilities with which the languages
$a_1^* \ldots a_n^*$ can be recognized.

There is also a lot of results about the number
of states needed for QFA to recognize different languages.
In some cases, it can be exponentially less than for deterministic or even
for probabilistic automata \cite{AF 98,K 98}.
In other cases, it can
be exponentially bigger than for deterministic
automata \cite{ANTV 98,N 99}.

A good survey about quantum automata is Gruska \cite{G 00}.
\end{subsection}

\section{Main results}

\subsection{Necessary condition}

First, we give the new condition which implies that the
language is not recognizable by a QFA.
Similarly to the previous condition (Theorems \ref{T13}), 
it can be formulated as a condition about the minimal deterministic
automaton of a language.

\begin{theorem}
\label{C2}
Let $L$ be a language. Assume that there are
words $x$, $y$, $z_1$, $z_2$ such that its minimal automaton $M$ 
contains states $q_1$, $q_2$, $q_3$ satisfying:

1. $q_2\neq q_3$,

2. if M starts in the state $q_1$ and reads $x$, it passes to $q_2$,

3. if M starts in the state $q_2$ and reads $x$, it passes to $q_2$,

4. if M starts in the state $q_1$ and reads $y$, it passes to $q_3$,

5. if M starts in the state $q_3$ and reads $y$, it passes to $q_3$,

6. for all words $t\in (x|y)^*$ there exists a word $t_1\in (x|y)^*$ such that
if M starts in the state $q_2$ and reads $tt_1$, it passes to $q_2$,

7. for all words $t\in (x|y)^*$ there exists a word $t_1\in (x|y)^*$ such that
if M starts in the state $q_3$ and reads $tt_1$, it passes to $q_3$,

8. if M starts in the state $q_2$ and reads $z_1$, it passes to
an accepting state,

9. if M starts in the state $q_2$ and reads $z_2$, it passes to
a rejecting state,

10. if M starts in the state $q_3$ and reads $z_1$, it passes to
a rejecting state,

11. if M starts in the state $q_3$ and reads $z_2$, it passes to
an accepting state.
 \\
Then $L$ cannot be recognized by a 1-way QFA. 
\end{theorem}

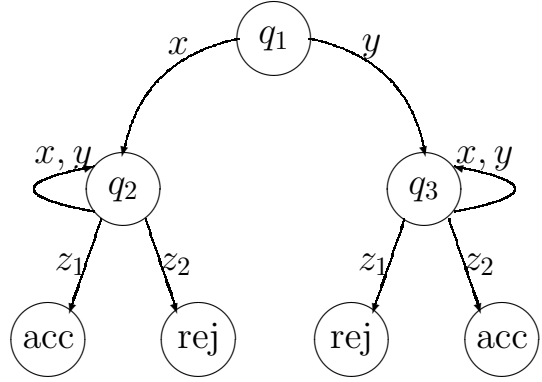
\begin{figure}[htb]
  \centering
  \begin{Large}
\unitlength 1.00mm
\linethickness{0.4pt}
\begin{picture}(80.00,50.00)
\put(20.00,25.00){\circle{10.00}}
\put(20.00,25.00){\makebox(0,0)[cc]{$q_2$}}
\put(60.00,25.00){\circle{10.00}}
\put(60.00,25.00){\makebox(0,0)[cc]{$q_3$}}
\put(68.00,29.00){\makebox(0,0)[cc]{$x,y$}}
\put(64.00,28.00){\vector(-4,1){0.2}}
\bezier{132}(64.00,22.00)(80.00,25.00)(64.00,28.00)
\put(12.00,29.00){\makebox(0,0)[cc]{$x,y$}}
\put(16.00,28.00){\vector(4,1){0.2}}
\bezier{132}(16.00,22.00)(0.00,25.00)(16.00,28.00)
\put(40.00,45.00){\circle{10.00}}
\put(20.00,30.00){\vector(-1,-4){0.2}}
\bezier{104}(35.33,45.00)(22.33,42.33)(20.00,30.00)
\put(60.00,30.00){\vector(1,-4){0.2}}
\bezier{104}(44.67,45.00)(57.67,42.33)(60.00,30.00)
\put(53.00,44.00){\makebox(0,0)[cc]{$y$}}
\put(27.33,44.00){\makebox(0,0)[cc]{$x$}}
\put(40.00,45.00){\makebox(0,0)[cc]{$q_1$}}
\put(10.00,5.00){\circle{10.00}}
\put(13.00,9.00){\vector(-1,-3){0.2}}
\multiput(17.00,21.00)(-0.12,-0.35){34}{\line(0,-1){0.35}}
\put(10.00,5.00){\makebox(0,0)[cc]{acc}}
\put(30.00,5.00){\circle{10.00}}
\put(27.00,9.00){\vector(1,-3){0.2}}
\multiput(23.00,21.00)(0.12,-0.35){34}{\line(0,-1){0.35}}
\put(30.00,5.00){\makebox(0,0)[cc]{rej}}
\put(13.00,15.00){\makebox(0,0)[cc]{$z_1$}}
\put(27.00,15.00){\makebox(0,0)[cc]{$z_2$}}
\put(50.33,5.00){\circle{10.00}}
\put(53.33,9.00){\vector(-1,-3){0.2}}
\multiput(57.33,21.00)(-0.12,-0.35){34}{\line(0,-1){0.35}}
\put(50.33,5.00){\makebox(0,0)[cc]{rej}}
\put(70.33,5.00){\circle{10.00}}
\put(67.33,9.00){\vector(1,-3){0.2}}
\multiput(63.33,21.00)(0.12,-0.35){34}{\line(0,-1){0.35}}
\put(70.33,5.00){\makebox(0,0)[cc]{acc}}
\put(53.33,15.00){\makebox(0,0)[cc]{$z_1$}}
\put(67.33,15.00){\makebox(0,0)[cc]{$z_2$}}
\end{picture}
  \end{Large}
  \caption{Conditions of theorem \ref{C2},
conditions 6 and 7 are shown symbolically}
  \label{Bilde2}
\end{figure}

\begin{proof}
We use a lemma from \cite{BV 97}.

\begin{lemma}
\label{LBV}
If $\psi$ and $\phi$ are two quantum states and
$\|\psi-\phi\|<\varepsilon$ then the total variational distance
between the probability distributions generated by the same
measurement on $\psi$ and $\phi$ is at most\footnote{The 
lemma in \cite{BV 97} has $4\varepsilon$ but 
it can be improved to $2\varepsilon$.} $2\varepsilon$.
\end{lemma}

We also use a lemma from \cite{AF 98}.

\begin{lemma}
\label{LemmaAF}
Let $x\in \Sigma^{+}$.
There are subspaces $E_1$, $E_2$ such that $E_{non}=E_1\oplus E_2$ and
\begin{enumerate}
\item[(i)]
If $\psi\in E_1$, then $V'_x(\psi)\in E_1$ and $\| V'_x(\psi)\|=\|\psi\|$,
\item[(ii)]
If $\psi\in E_2$, then $\| V'_{x^k}(\psi)\|\rightarrow 0$ when
$k\rightarrow\infty$.
\end{enumerate}
\end{lemma}

Lemma \ref{LemmaAF} can be viewed as a quantum counterpart of
the {\em classification of states for Markov chains} \cite{KS 76}.
The classification of states divides the states of a Markov
chain into {\em ergodic} sets and {\em transient} sets.
If the Markov chain is in an ergodic set, it never leaves it.
If it is in a transient set, it leaves it with probability
$1-\epsilon$ for an arbitrary $\epsilon>0$ after
sufficiently many steps.

In the quantum case, $E_1$ is the counterpart of an ergodic set:
if the quantum random process defined by repeated reading of $x$
is in a state $\psi\in E_1$, it stays in $E_1$.
$E_2$ is a counterpart of a transient set:
if the state is $\psi\in E_2$, $E_2$ is left (for an accepting
or rejecting state) with probability arbitrarily close to 1
after sufficiently many $x$'s.

The next Lemma is our generalization of Lemma \ref{LemmaAF}
for the case of two different words $x$ and $y$.

\begin{lemma}
\label{Lemma2}
Let $x,y\in \Sigma^{+}$.
There are subspaces $E_1$, $E_2$ such that $E_{non}=E_1\oplus E_2$ and
\begin{enumerate}
\item[(i)]
If $\psi\in E_1$, then $V'_x(\psi)\in E_1$ and $V'_y(\psi)\in E_1$
and $\|V'_x(\psi)\|=\|\psi\|$ and $\|V'_y(\psi)\|=\|\psi\|$,
\item[(ii)]
If $\psi\in E_2$, then for any $\epsilon >0$,
there exists a word $t\in (x|y)^*$ such that
$\| V'_t(\psi)\|<\epsilon$.
\end{enumerate}
\end{lemma}

\noindent
{\em Proof.}
We use $E_1^z$ to denote the space $E_1$ from Lemma \ref{LemmaAF}
for a word $z$.
We define $E_1=\bigcap\limits _{z\in (x|y)^*} E_1^z$.
$E_2$ consists of all vectors in $E_{non}$ orthogonal to $E_1$.
Next, we check that both (i) and (ii) are true.\\

(i)
It is easy to see that, for all $t\in (x|y)^*$,
$\|V'_t(\psi)\|=\|\psi\|$ due to $\psi\in E^t_1$.

We also need to prove that $V'_x(\psi)\in E_1$ and $V'_y(\psi)\in E_1$.
For a contradiction, assume there are $\psi\in E_1$ and $t_1\in (x|y)^*$
such that $V'_{t_1}(\psi)\notin E_1$.
Then, by definition of $E_1$, there also exists $t_2\in (x|y)^*$
such that $V'_{t_1}(\psi)$ does not belong to $E_1^{t_2}$.
Lemma \ref{LemmaAF} implies that the norm of $V'_{t_1}(\psi)$ can be
decreased by repeated applications of $V'_{t_2}$.
A contradiction with $\|V'_t(\psi)\|=\|\psi\|$ for all $t$.\\

(ii) Clearly, if $\psi$ belongs to $E_2$ then for all $t\in (x|y)^*$
the superposition $V'_t(\psi)$ also belongs to $E_2$ because
$V_x$ and $V_y$ are unitary and map $E_1$ to itself (and, therefore,
any vector orthogonal to $E_1$ is mapped to a vector
orthogonal to $E_1$).

$\|V'_t(\psi)\|$ does not increase if we extend the word $t$ to the right
and it is bounded from below by 0.
Hence, for any fixed $\epsilon$ we can find a $t\in (x|y)^*$ such that
$$\|V'_t(\psi)\|-\|V'_{tw}(\psi)\|<\epsilon$$
for all $w\in (x|y)^*$.
We define a sequence of such words $t_1, t_2, t_3, \ldots$ for
$\epsilon, \frac{\epsilon}{2}, \frac{\epsilon}{4},\ldots$.
$V'_{t_1}(\psi), V'_{t_2}(\psi), V'_{t_3}(\psi),\ldots$
is a bounded sequence in a finite dimensional space. 
Therefore, it has a limit point $\psi '$.
We will show that $\psi'=0$.

First, notice that $\psi'\in E_2$ because it is a
limit of a subsequence of 
$V'_{t_1}(\psi), V'_{t_2}(\psi), V'_{t_3}(\psi),\ldots$
and all $V'_{t_i}(\psi)$ belong to $E_2$.
Therefore, if $\psi'\neq 0$ then
$\psi'$ has nonzero $E^z_2$ component
for some $z\in (x|y)^{*}$.
Reading sufficiently many $z$ would decrease this component,
decreasing the norm of $\psi'$.

This contradicts the fact that, for any $w$, 
$\|\psi'\|=\|V'_{w}(\psi')\|$
(since $\|\psi'\|-\|V'_{w}(\psi')\|$ is less than any $\epsilon>0$
which is true because $\psi'$ is the limit of  
$V'_{t_1}(\psi), V'_{t_2}(\psi), V'_{t_3}(\psi),\ldots$).

Therefore, $\psi'=0$. This completes the proof of lemma.
\qed 

Let $L$ be a language such that its minimal automaton $M$
contains the "forbidden construction" and $M_q$ be a QFA.
We show that $M_q$ does not recognize $L$.

Let $w$ be a word after reading which $M$
is in the state $q_1$.
Let $\psi_{w}=\psi^1_{w}+\psi^2_{w}$,
$\psi^1_{w}\in E_1$, $\psi^2_{w}\in E_2$.
We find a word $a\in (x|y)^*$ such that after reading $xa$ 
$M$ is in the state $q_2$ and the norm of 
$\psi^2_{wxa}=V'_a(\psi^2_{wx})$ is at most 
some fixed $\epsilon>0$.
(Such word exists due to Lemma \ref{Lemma2} and
conditions 6 and 7.)
We also find a word $b$ such that $\|\psi^2_{wyb}\|\leq \epsilon$.

Because of unitarity of $V'_x$ and $V'_y$ on $E_1$ 
(part (i) of Lemma \ref{Lemma2}),
there exist integers $i$ and $j$ such that
$\|\psi^1_{w(xa)^i}-\psi^1_{w}\|\leq\epsilon$
and
$\|\psi^1_{w(yb)^j}-\psi^1_{w}\|\leq\epsilon$.


Let $p$ be the probability of $M_q$ accepting while reading $\kappa w$.
Let $p_1$ be the probability of accepting while reading $(xa)^i$
with a starting state $\psi_w$,
$p_2$ be the probability of accepting while reading $(yb)^j$
with a starting state $\psi_w$ and $p_3$, $p_4$ be
the probabilities of accepting while reading $z_1\$$ and $z_2\$$
with a starting state $\psi^1_w$.

Let us consider four words
$\kappa w (xa)^iz_1\$$,
$\kappa w (xa)^iz_2\$$,
$\kappa w (yb)^jz_1\$$,
$\kappa w (yb)^jz_2\$$.

\begin{lemma}
$M_q$ accepts $\kappa w (xa)^iz_1\$$
with probability at least $p+p_1+p_3-4\epsilon$
and at most $p+p_1+p_3+4\epsilon$.
\end{lemma}

\noindent
{\em Proof.}
The probability of accepting while reading $\kappa w$ is $p$.
After that, $M_q$ is in the state $\psi_w$ and reading $(xa)^i$
in this state causes it to accept with probability $p_1$.

The remaining state is $\psi_{w(xa)^i}=\psi^1_{w(xa)^i}+\psi^2_{w(xa)^i}$.
If it was $\psi^1_w$, the probability of accepting while
reading the rest of the word ($z_1\$$) would be exactly $p_3$.
It is not quite $\psi^1_w$ but it is close to $\psi^1_w$.
Namely, we have 
\[ \|\psi_{w(xa)^i}-\psi^1_w\| \leq \|\psi^2_{w(xa)^i}\|
+\|\psi^1_{w(xa)^i}-\psi^1_w\| \leq \epsilon+\epsilon =2\epsilon.\]
By Lemma \ref{LBV}, this means that the probability of accepting
during $z_1\$$ is between $p_3-4\epsilon$ and $p_3+4\epsilon$.
\qed

Similarly, on the second word $M_q$ accepts with probability between 
$p+p_1+p_4-4\epsilon$ and $p+p_1+p_4+4\epsilon$.
On the third word $M_q$ accepts with probability between 
$p+p_2+p_3-4\epsilon$ and $p+p_2+p_3+4\epsilon$.
On the fourth word $M_q$ accepts with probability 
$p+p_2+p_4-4\epsilon$ and $p+p_2+p_4+4\epsilon$.

This means that the sum of accepting probabilities of two words that belong to $L$
(the first and the fourth words)
differs from the sum of accepting probabilities of two words that do not
belong to $L$ (the second and the third) by at most $16\epsilon$.
Hence, the probability of correct answer of $M_q$ 
on one of these words is at most $\frac{1}{2}+4\epsilon$.
Since such 4 words can be constructed for 
arbitrarily small $\epsilon$,
this means that $M_q$ does not recognize $L$.
\qed
\end{proof}

\subsection{Necessary and sufficient condition}
      
\begin{figure}[htb]
  \centering
  \begin{Large}
\unitlength 1.00mm
\linethickness{0.4pt}
\begin{picture}(100.00,15.00)
\put(20.00,10.00){\circle{10.00}}
\put(50.00,10.00){\circle{10.00}}
\put(20.00,10.00){\makebox(0,0)[cc]{$q_1$}}
\put(50.00,10.00){\makebox(0,0)[cc]{$q_2$}}
\put(46.00,7.00){\vector(3,1){0.2}}
\bezier{92}(24.00,7.00)(35.00,3.00)(46.00,7.00)
\put(35.00,7.00){\makebox(0,0)[cc]{$x$}}
\put(58.00,14.00){\makebox(0,0)[cc]{$x$}}
\put(80.00,10.00){\circle{10.00}}
\put(54.00,13.00){\vector(-4,1){0.2}}
\bezier{132}(54.00,7.00)(70.00,10.00)(54.00,13.00)
\put(76.00,7.00){\vector(3,1){0.2}}
\bezier{92}(54.00,7.00)(65.00,3.00)(76.00,7.00)
\put(80.00,10.00){\makebox(0,0)[cc]{$q_3$}}
\put(65.00,7.00){\makebox(0,0)[cc]{$y$}}
\put(88.00,14.00){\makebox(0,0)[cc]{$y$}}
\put(84.00,13.00){\vector(-4,1){0.2}}
\bezier{132}(84.00,7.00)(100.00,10.00)(84.00,13.00)
\end{picture}
  \end{Large}
  \caption{Conditions of theorem \ref{T14}}
  \label{Bilde10}
\end{figure}
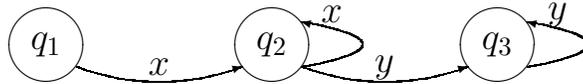

For languages whose minimal automaton does not contain
the construction of Figure \ref{Bilde10}, this condition
(together with Theorem \ref{T13}) is necessary and sufficient.

\begin{theorem}
\label{T14}
Let $U$ be the class of languages whose minimal automaton does
not contain "two cycles in a row" (Fig. \ref{Bilde10}).
A language that belongs to $U$ can be recognized by a 1-way QFA
if and only if its minimal deterministic
automaton does not contain the "forbidden construction" from
Theorem \ref{T13} and the "forbidden construction" from Theorem \ref{C2}.
\end{theorem}

\begin{proof}
Let $M$ be the minimal deterministic automaton of a language $L$.
If it contains at least one of "forbidden constructions" of Theorems
\ref{T13} and \ref{C2}, then $L$
cannot be recognized by a 1-way QFA.
We now show that, if $M$ does not contain any of the two
``forbidden constructions'' and does not contain
``two cycles in a row'' construction then $L$ can be
recognized by a QFA.

Let $q_0$ be the starting state of $M$ and $V$ be the
transition function of the automaton $M$.
$V(q, x)$ denotes the state to which $M$ goes if
it reads the word $x$ in the state $q$.

We will construct a QFA for $L$ by splitting $M$ into
pieces $A$, $B_1$, $\ldots$, $B_n$, constructing
a reversible finite automaton for each of those pieces
and then combining these reversible automata.

Let $B$ be the set of all states $q$ such that
after reading any word in $q$, 
there exists a word such that $M$ passes back to the state $q$.
We split $B$ into connected components $B_1$, $B_2$, $\ldots$, $B_n$.
Two different states $q_i$ and $q_j$ belong to the
same $B_k$ iff $q_i$ is reachable from $q_j$ and $q_j$ is
reachable from $q_i$.
Let $A$ be the set of all remaining states, i. e., the states that
do not belong to $B=B_1\cup B_2\cup \ldots \cup B_n$.

\begin{figure}[htb]
  \centering
  \begin{Large}
  
  \unitlength 1.00mm
  \linethickness{0.4pt}

  \begin{picture}(80.00,45.00)

  \put(40.00,40.00){\circle{10.00}}
  \put(40.00,40.00){\makebox(0,0)[cc]{$A$}} 
  \put(10.00,10.00){\circle{10.00}}
  \put(10.00,10.00){\makebox(0,0)[cc]{$B_1$}}
  \put(30.00,10.00){\circle{10.00}}
  \put(30.00,10.00){\makebox(0,0)[cc]{$B_2$}}
  \put(50.00,10.00){\makebox(0,0)[cc]{$\ldots$}}
  \put(70.00,10.00){\circle{10.00}}
  \put(70.00,10.00){\makebox(0,0)[cc]{$B_n$}}
 
  \put(13.53,13.53){\vector(-1,-1){0.2}}
  \put(36.47,36.47){\line(-1,-1){23}}

  \put(31.58,14.74){\vector(-1,-3){0.2}}
  \put(38.42,35.26){\line(-1,-3){7}}

  \put(66.47,13.53){\vector(1,-1){0.2}}
  \put(43.53,36.47){\line(1,-1){23}}

  \end{picture}
  \end{Large}

  \caption{Division of $M$}
  \label{BildePedeja}
\end{figure}
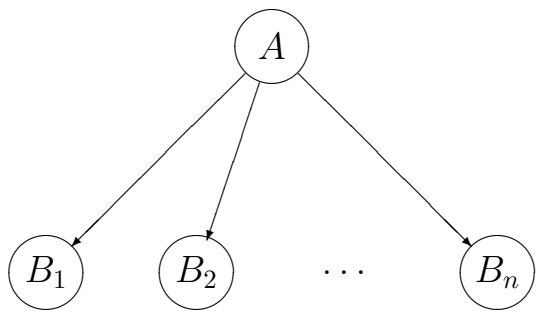


\begin{lemma}
\label{newlemma1}
For every letter $a$ and every state $q$ of $B_i$,
there is exactly one $q'\in B_i$ such that reading $a$ in $q'$ leads to $q$,
i. e., every letter induces a permutation of states in $B_i$.
\end{lemma}

\begin{proof}
Let $q$ be a state in $B_i$ and $a\in\Sigma$.
For a contradiction, assume that there are two states $q'\in B_i$ 
and $q''\in B_i$ such that $V(q', a)=V(q'', a)=q$.

Then, there exist words $x'$ and $x''$ such that $V(q, x'')=q''$
and $V(q'', x')=q'$. (This is true because every state in $B_i$ is reachable
from every other state in $B_i$.)

However, this means that $B_i$ contains the ``forbidden construction''
of Theorem \ref{T13}, with $q_1=q'$, $q_2=q''$, $x=ax''$ and $y=x'$.
A contradiction.
\qed
\end{proof}

Such automata $B_i$ are called {\em permutation automata}.

\hspace{0.25\textwidth}
\begin{figure}[htb]
  \centering
  \begin{Large}
\unitlength 1.00mm
\linethickness{0.4pt}
\begin{picture}(70.00,15.00)
\put(20.00,10.00){\circle{10.00}}
\put(50.00,10.00){\circle{10.00}}
\put(20.00,10.00){\makebox(0,0)[cc]{$q_1$}}
\put(50.00,10.00){\makebox(0,0)[cc]{$q_2$}}
\put(54.00,7.00){\vector(-4,-1){0.2}}
\bezier{132}(54.00,13.00)(70.00,10.00)(54.00,7.00)
\put(46.00,7.00){\vector(3,1){0.2}}
\bezier{92}(24.00,7.00)(35.00,3.00)(46.00,7.00)
\put(35.00,7.00){\makebox(0,0)[cc]{$x$}}
\put(58.00,14.00){\makebox(0,0)[cc]{$x$}}
\end{picture}
  \end{Large}
  \caption{Conditions of Lemma \ref{newlemma2}}
  \label{Bilde1}
\end{figure}
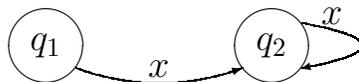
\hspace{0.25\textwidth}

\begin{lemma}
\label{newlemma2}
$A$ does not contain a fragment of the form shown in Fig. \ref{Bilde1}
with $q_1$ and $q_2$ being two different states of $A$ and
$x\in\Sigma^+$.
\end{lemma}

\begin{proof}
For a contradiction, assume that $A$ contains such a fragment.

By definition of $A$ and $B$,
$q_2\in A$ implies that there is a word $z$ such that reading $z$
in $q_2$ leads to a state $q_2'$ and $q_2$ is not reachable
from $q_2'$. 

Consider the states $V(q_2, z)$, $V(q_2, z^2)$, $\ldots$.
$M$ has a finite number of states. 
Therefore, there must be $j$ and $k$ such that 
$V(q_2, z^j)=V(q_2, z^{j+k})$.
Notice that this implies $V(q_2, z^{j'})=V(q_2, z^{j'+k})$
for all $j'\geq j$.

Let $i$ be the smallest number such that $i\geq j$ and $i$ is 
divisible by $k$.
Define $q_3=V(q_2, z^i)$.
Then, $V(q_2, z^i)=V(q_2, z^{i+k})=V(q_2, z^{i+2k})=\ldots=V(q_2, z^{2i})$
(because $i$ is divisible by $k$), i. e., $q_3=V(q_3, z^i)$.

We have shown that 
$q_1$, $q_2$, $q_3$ form a ``two cycles in a row'' construction
with $y=z^i$. 
A contradiction.
\qed
\end{proof}

By a theorem from \cite{AF 98}, any language recognizable by
a deterministic automaton which does not contain the construction
of Fig. \ref{Bilde1} is recognizable by a reversible finite automaton(RFA).
(A reversible finite automaton is a deterministic
automaton in which, for every state $q$ and letter $a$, 
there is at most one state $q'$ such that reading $a$ in $q'$ leads
to $q$.)

Any reversible automaton is a special case of a quantum automaton.
(If, for every state $q$ and every letter $a$, there is one $q'$
such that reading $a$ leads to $q$, the letter $a$ induces a permutation
on states of automaton and the corresponding transformation
of a quantum automaton is clearly unitary.)

Therefore, the language recognized by $A$ is recognized by a QFA as well.
Also, permutation automata $B_1$, $\ldots$, $B_n$ are 
special cases of reversible automata.
Therefore, they can be replaced by equivalent QFAs.
We will construct a QFA for $L$ by combining those QFAs.

However, before that, we must solve one problem. 
Even if the state of $M$ after reading a word $x$ is in $B_i$,
the starting state of $M$ can be in $A$.
If we want to use the permutation automaton for $B_i$ to
recognize a part of $L$, we must define one of states in $B_i$
as the starting state.
The next two lemmas show that this is possible.

\begin{lemma}
\label{newlemma3}
If the minimal automaton $M$ contains the construction of Fig. \ref{Co5},
the states $q_2$
and $q_3$ cannot be in the same $B_i$.
\end{lemma}

\begin{proof}
Let us suppose the opposite.

$q_2$ and $q_3$ are different states of the minimal deterministic
automaton.
Therefore, there exists a word $a$ such that
$V(q_2,a)$ is an accepting state (or a rejecting state) and
$V(q_3,a)$ is a rejecting state (or an accepting state).

Also, there is no word $l$ such that
$V(q_2,l)$ is a rejecting state (or an accepting state) and
$V(q_3,l)$ is an accepting state (or a rejecting state) because,
otherwise, $M$ would contain the construction of Theorem \ref{C2}.

We denote $V(q_2,a)$ by $q_{acc}$ and $V(q_3,a)$ by $q_{rej}$.
There exists a word $b$ such that $V(q_{rej},b)=q_{acc}$
(because all states in $B_i$ can be reached from one another).
Moreover,
the states $V(q_2,ab)$ and $V(q_3,ab)$ are accepting states
($V(q_3, ab)$ is accepting because $V(q_3, ab)=V(q_{rej}, b)=q_{acc}$
and $V(q_2, ab)$ is accepting because, if it was rejecting,
$q_1$, $q_2$ and $q_3$ would form the construction of Theorem \ref{C2}
with $a$ and $ab$ as $z_1$ and $z_2$.).
Similarly, the states $V(q_2,abb)$ and $V(q_3,abb)$ are accepting states,
the states $V(q_2,abbb)$ and $V(q_3,abbb)$ are accepting states and so on.
However, there exists $k$ such that $V(q_3,ab^k)=V(q_3, a)=q_{rej}$
(because $B_i$ is a permutation automaton and, therefore,
it must return to the starting state after some number of $b$'s).
This gives us the contradiction.
\qed
\end{proof}

\hspace{0.25\textwidth}
\begin{figure}[htb]
  \centering
  \begin{Large}
\unitlength 1.00mm
\linethickness{0.4pt}
\begin{picture}(75.00,31.00)
\put(15.00,6.00){\circle{10.00}}
\put(15.00,6.00){\makebox(0,0)[cc]{$q_2$}}
\put(55.00,6.00){\circle{10.00}}
\put(55.00,6.00){\makebox(0,0)[cc]{$q_3$}}
\put(63.00,10.00){\makebox(0,0)[cc]{$y$}}
\put(59.00,9.00){\vector(-4,1){0.2}}
\bezier{132}(59.00,3.00)(75.00,6.00)(59.00,9.00)
\put(7.00,10.00){\makebox(0,0)[cc]{$x$}}
\put(11.00,9.00){\vector(4,1){0.2}}
\bezier{132}(11.00,3.00)(-5.00,6.00)(11.00,9.00)
\put(35.00,26.00){\circle{10.00}}
\put(15.00,11.00){\vector(-1,-4){0.2}}
\bezier{104}(30.33,26.00)(17.33,23.33)(15.00,11.00)
\put(55.00,11.00){\vector(1,-4){0.2}}
\bezier{104}(39.67,26.00)(52.67,23.33)(55.00,11.00)
\put(48.00,25.00){\makebox(0,0)[cc]{$y$}}
\put(22.33,25.00){\makebox(0,0)[cc]{$x$}}
\put(35.00,26.00){\makebox(0,0)[cc]{$q_1$}}
\end{picture}
  \end{Large}
  \caption{}
  \label{Co5}
\end{figure}
\hspace{0.25\textwidth}


\begin{lemma}
\label{newlemma4}
For each part $B_i$, there is
a state $q$ such that if $V(q_0,x)$ belongs to $B_i$ then
$V(q_0,x)=V(q,x)$. 
\end{lemma}

\begin{proof}
Let $V(q_0, x)\in B_i$.
Then, there is a unique $q\in B_i$ such that
$V(q_0, x)=V(q, x)$. (This is true because $B_i$
is a permutation automaton and every state
has a unique preceding state.)
We must show that this state $q$ does not depend on the word $x$.

Assume this is not true.
Then, there are words $x$ and $y$ such that $V(q_0, x)=V(q_2, x)$
and $V(q_0, y)=V(q_3, y)$ and $q_2\neq q_3$
(and $V(q_2, x)$, $q_2$, $V(q_3, y)$, $q_3$ are all in $B_i$).

Let $q_4=V(q_2, x)$ and $q_5=V(q_3, y)$.
Then, there exist $j\in\bbbn$ and $k\in\bbbn$ such that
$V(q_4, x^j)=q_4$ and $V(q_5, y^k)=q_5$.
(Again, we are using the fact that $B_i$ is a 
permutation automaton, and, therefore, if it reads the same
word many times, it returns to the same state at some
point.)

This implies $V(q_4, x^{j-1})=q_2$ and $V(q_5, y^{k-1})=q_3$
(because $V(q_2, x)=q_4$ and $V(q_3, y)=q_5$ and, in a permutation automaton,
$q$ such that $V(q, x)=q_4$ must be unique).
Therefore, $V(q_0, x^j)=V(q_4, x^{j-1})=q_2$ and
$V(q_2, x^j)=V(q_4, x^{j-1})=q_2$.
Similarly, $V(q_0, y^k)=V(q_3, y^k)=q_3$.

This means that $B_i$ contains the states $q_2$
and $q_3$ from the construction shown in Figure \ref{Co5} (with $x^j$ and
$y^k$ instead of $x$ and $y$ and $q_0$ instead of $q_1$).
By Lemma \ref{newlemma3}, this is impossible.
A contradiction.
\qed
\end{proof}

We denote these states as $q_i$.
Let $B'_i$ be the automaton $B_i$ with $q_i$ as the starting state.
Let $L_i$ be the language recognized by $B'_i$.

\begin{lemma}
\label{newlemma5}
For any $i, j\in\{1, \ldots, n\}$,
either $L_i\subseteq L_j$ or $L_j\subseteq L_i$.
\end{lemma}

\begin{proof}
Let $x$ and $y$ be such that $V(q_0, x)=q_i$ and $V(q_0, y)=q_j$.
($x$ and $y$ exist because, otherwise, $q_i$ or $q_j$
would be unreachable from the starting state $q_0$.)
By Lemma \ref{newlemma4}, $V(q_i, x)=q_i$ and $V(q_j, y)=q_j$.

For a contradiction, assume that neither $L_i\subseteq L_j$ 
nor $L_j\subseteq L_i$ is true. 
Then, there are words $z_1\in L_i-L_j$ and $z_2\in L_j-L_i$
and we get the ``forbidden construction'' of Theorem \ref{C2}.
\qed
\end{proof}

Let $a_i$ denote the number of $j\in\{1, \ldots, n\}$ such that 
$L_j\subseteq L_i$.
%
$A'$ denotes the corresponding reversible automaton for the
automaton $A$ with one modification:
when the automaton $M$ passes to a state of
$B_1\cup B_2 \cup \ldots \cup B_n$
$A'$ accepts with probability $\frac{n-a_i}{n+1}$ and rejects
with probability $\frac{a_i+1}{n+1}$.

Next, we define a QFA recognizing the language $L$:
it works as $A'$ with probability $p=\frac{n+1}{2n+1}$
(with amplitude $\sqrt{\frac{n+1}{2n+1}}$)
and as $B_i$ with probability $\frac{1}{2n+1}$
(with amplitude $\frac{1}{\sqrt{2n+1}}$) for each $i$.

{\it Case 1.} $V(q_0,x)\in A$.
The QFA recognizes $x$ with probability $p$.

{\it Case 2.} $V(q_0,x)\in B_i$ and $x\in L$.
The automaton $A'$ accepts with probability $\frac{n-a_i}{n+1}$.
Moreover, $x$ is accepted by at least $a_i$ automata
from $B_1', B_2',\ldots ,B_n'$. This means that the total probability of
accepting is at least
$$\frac{n-a_i}{n+1}\cdot \frac{n+1}{2n+1}+\frac{a_i+1}{2n+1}=
\frac{n+1}{2n+1}=p.$$

{\it Case 3.} $V(q_0,x)\in B_i$ and $x\notin L$.
Similarly to the previous case, the total probability of
rejecting is at least $p$.
\qed
\end{proof}

\section{Non-closure under union}

\subsection{Non-closure result}
                                 
In particular, Theorem \ref{C2} implies that the class
of languages recognized by QFAs is not closed under union.

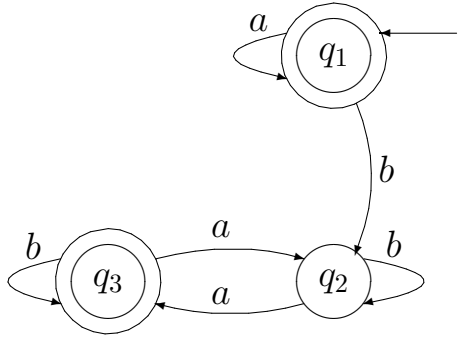
\begin{figure}[htb]
  \centering
  \begin{Large}
\special{em:linewidth 0.4pt}
\unitlength 1mm
\linethickness{0.4pt}
\begin{picture}(70.00,47.00)
\emline{50.00}{47.00}{1}{51.59}{46.82}{2}
\emline{51.59}{46.82}{3}{53.09}{46.28}{4}
\emline{53.09}{46.28}{5}{54.43}{45.42}{6}
\emline{54.43}{45.42}{7}{55.54}{44.27}{8}
\emline{55.54}{44.27}{9}{56.37}{42.91}{10}
\emline{56.37}{42.91}{11}{56.86}{41.39}{12}
\emline{56.86}{41.39}{13}{57.00}{39.80}{14}
\emline{57.00}{39.80}{15}{56.77}{38.22}{16}
\emline{56.77}{38.22}{17}{56.19}{36.73}{18}
\emline{56.19}{36.73}{19}{55.29}{35.42}{20}
\emline{55.29}{35.42}{21}{54.11}{34.34}{22}
\emline{54.11}{34.34}{23}{52.72}{33.55}{24}
\emline{52.72}{33.55}{25}{51.19}{33.10}{26}
\emline{51.19}{33.10}{27}{49.60}{33.01}{28}
\emline{49.60}{33.01}{29}{48.03}{33.28}{30}
\emline{48.03}{33.28}{31}{46.56}{33.90}{32}
\emline{46.56}{33.90}{33}{45.27}{34.84}{34}
\emline{45.27}{34.84}{35}{44.22}{36.05}{36}
\emline{44.22}{36.05}{37}{43.48}{37.46}{38}
\emline{43.48}{37.46}{39}{43.07}{39.00}{40}
\emline{43.07}{39.00}{41}{43.03}{40.60}{42}
\emline{43.03}{40.60}{43}{43.34}{42.16}{44}
\emline{43.34}{42.16}{45}{44.01}{43.61}{46}
\emline{44.01}{43.61}{47}{44.98}{44.88}{48}
\emline{44.98}{44.88}{49}{46.22}{45.89}{50}
\emline{46.22}{45.89}{51}{47.65}{46.59}{52}
\emline{47.65}{46.59}{53}{50.00}{47.00}{54}
\put(50.00,40.00){\circle{10.00}}
\put(50.00,40.00){\makebox(0,0)[cc]{$q_1$}}
\emline{20.00}{17.00}{55}{21.59}{16.82}{56}
\emline{21.59}{16.82}{57}{23.09}{16.28}{58}
\emline{23.09}{16.28}{59}{24.43}{15.42}{60}
\emline{24.43}{15.42}{61}{25.54}{14.27}{62}
\emline{25.54}{14.27}{63}{26.37}{12.91}{64}
\emline{26.37}{12.91}{65}{26.86}{11.39}{66}
\emline{26.86}{11.39}{67}{27.00}{9.80}{68}
\emline{27.00}{9.80}{69}{26.77}{8.22}{70}
\emline{26.77}{8.22}{71}{26.19}{6.73}{72}
\emline{26.19}{6.73}{73}{25.29}{5.42}{74}
\emline{25.29}{5.42}{75}{24.11}{4.34}{76}
\emline{24.11}{4.34}{77}{22.72}{3.55}{78}
\emline{22.72}{3.55}{79}{21.19}{3.10}{80}
\emline{21.19}{3.10}{81}{19.60}{3.01}{82}
\emline{19.60}{3.01}{83}{18.03}{3.28}{84}
\emline{18.03}{3.28}{85}{16.56}{3.90}{86}
\emline{16.56}{3.90}{87}{15.27}{4.84}{88}
\emline{15.27}{4.84}{89}{14.22}{6.05}{90}
\emline{14.22}{6.05}{91}{13.48}{7.46}{92}
\emline{13.48}{7.46}{93}{13.07}{9.00}{94}
\emline{13.07}{9.00}{95}{13.03}{10.60}{96}
\emline{13.03}{10.60}{97}{13.34}{12.16}{98}
\emline{13.34}{12.16}{99}{14.01}{13.61}{100}
\emline{14.01}{13.61}{101}{14.98}{14.88}{102}
\emline{14.98}{14.88}{103}{16.22}{15.89}{104}
\emline{16.22}{15.89}{105}{17.65}{16.59}{106}
\emline{17.65}{16.59}{107}{20.00}{17.00}{108}
\put(20.00,10.00){\circle{10.00}}
\put(50.00,10.00){\circle{10.00}}
\put(26.33,7.00){\vector(-4,1){0.2}}
\emline{46.00}{7.00}{109}{43.58}{6.42}{110}
\emline{43.58}{6.42}{111}{41.15}{6.00}{112}
\emline{41.15}{6.00}{113}{38.70}{5.75}{114}
\emline{38.70}{5.75}{115}{33.79}{5.75}{116}
\emline{33.79}{5.75}{117}{31.31}{6.00}{118}
\emline{31.31}{6.00}{119}{28.83}{6.42}{120}
\emline{28.83}{6.42}{121}{26.33}{7.00}{122}
\put(46.00,13.00){\vector(4,-1){0.2}}
\emline{26.33}{13.00}{123}{28.97}{13.58}{124}
\emline{28.97}{13.58}{125}{31.56}{14.00}{126}
\emline{31.56}{14.00}{127}{34.10}{14.25}{128}
\emline{34.10}{14.25}{129}{39.02}{14.25}{130}
\emline{39.02}{14.25}{131}{41.40}{14.00}{132}
\emline{41.40}{14.00}{133}{43.72}{13.58}{134}
\emline{43.72}{13.58}{135}{46.00}{13.00}{136}
\put(20.00,10.00){\makebox(0,0)[cc]{$q_3$}}
\put(50.00,10.00){\makebox(0,0)[cc]{$q_2$}}
\put(54.00,7.00){\vector(-4,-1){0.2}}
\emline{54.00}{13.00}{137}{56.24}{12.55}{138}
\emline{56.24}{12.55}{139}{58.11}{12.09}{140}
\emline{58.11}{12.09}{141}{59.62}{11.64}{142}
\emline{59.62}{11.64}{143}{60.76}{11.18}{144}
\emline{60.76}{11.18}{145}{61.53}{10.73}{146}
\emline{61.53}{10.73}{147}{61.93}{10.27}{148}
\emline{61.93}{10.27}{149}{61.97}{9.82}{150}
\emline{61.97}{9.82}{151}{61.64}{9.36}{152}
\emline{61.64}{9.36}{153}{60.94}{8.91}{154}
\emline{60.94}{8.91}{155}{59.88}{8.45}{156}
\emline{59.88}{8.45}{157}{58.44}{8.00}{158}
\emline{58.44}{8.00}{159}{56.64}{7.55}{160}
\emline{56.64}{7.55}{161}{54.00}{7.00}{162}
\put(13.67,7.00){\vector(4,-1){0.2}}
\emline{13.67}{13.00}{163}{11.46}{12.48}{164}
\emline{11.46}{12.48}{165}{9.67}{11.97}{166}
\emline{9.67}{11.97}{167}{8.30}{11.45}{168}
\emline{8.30}{11.45}{169}{7.34}{10.93}{170}
\emline{7.34}{10.93}{171}{6.80}{10.41}{172}
\emline{6.80}{10.41}{173}{6.67}{9.90}{174}
\emline{6.67}{9.90}{175}{6.97}{9.38}{176}
\emline{6.97}{9.38}{177}{7.67}{8.86}{178}
\emline{7.67}{8.86}{179}{8.80}{8.34}{180}
\emline{8.80}{8.34}{181}{10.34}{7.83}{182}
\emline{10.34}{7.83}{183}{13.67}{7.00}{184}
\put(53.00,14.00){\vector(-1,-3){0.2}}
\emline{53.00}{33.67}{185}{53.84}{31.36}{186}
\emline{53.84}{31.36}{187}{54.45}{29.04}{188}
\emline{54.45}{29.04}{189}{54.84}{26.72}{190}
\emline{54.84}{26.72}{191}{55.00}{24.38}{192}
\emline{55.00}{24.38}{193}{54.93}{22.04}{194}
\emline{54.93}{22.04}{195}{54.63}{19.69}{196}
\emline{54.63}{19.69}{197}{54.11}{17.32}{198}
\emline{54.11}{17.32}{199}{53.00}{14.00}{200}
\put(57.00,25.00){\makebox(0,0)[cc]{$b$}}
\put(58.00,15.00){\makebox(0,0)[cc]{$b$}}
\put(35.00,17.00){\makebox(0,0)[cc]{$a$}}
\put(35.00,8.00){\makebox(0,0)[cc]{$a$}}
\put(10.00,14.67){\makebox(0,0)[cc]{$b$}}
\put(56.33,43.00){\vector(-1,0){0.2}}
\emline{66.33}{43.00}{201}{56.33}{43.00}{202}
\put(43.67,37.00){\vector(4,-1){0.2}}
\emline{43.67}{43.00}{203}{41.46}{42.48}{204}
\emline{41.46}{42.48}{205}{39.67}{41.97}{206}
\emline{39.67}{41.97}{207}{38.30}{41.45}{208}
\emline{38.30}{41.45}{209}{37.34}{40.93}{210}
\emline{37.34}{40.93}{211}{36.80}{40.41}{212}
\emline{36.80}{40.41}{213}{36.67}{39.90}{214}
\emline{36.67}{39.90}{215}{36.97}{39.38}{216}
\emline{36.97}{39.38}{217}{37.67}{38.86}{218}
\emline{37.67}{38.86}{219}{38.80}{38.34}{220}
\emline{38.80}{38.34}{221}{40.34}{37.83}{222}
\emline{40.34}{37.83}{223}{43.67}{37.00}{224}
\put(40.00,44.33){\makebox(0,0)[cc]{$a$}}
\end{picture}

  \end{Large}
  \caption{Automaton $G_1$}
  \label{Bilde2a}
\end{figure}

Let $L_1$ be the language consisting of
all words that start with \underline{any} number of
letters $a$ and after first letter $b$ (if there is one) there
is an odd number of letters $a$.
Its minimal automaton $G_1$ is shown in Fig.\ref{Bilde2a}.

This language satisfies the conditions of Theorem \ref{C2}.
($q_1$, $q_2$ and $q_3$ of Theorem \ref{C2} are just $q_1$, $q_2$ and $q_3$ 
of $G_1$. $x$, $y$, $z_1$ and $z_2$ are $b$, $aba$, $ab$ and $b$.) 
Hence, it cannot be recognized by a QFA.

Consider 2 other languages $L_2$ and $L_3$ defined as follows.

$L_2$ consists of all words which start with an \underline{even} number of
letters $a$ and after first letter $b$ (if there is one) there
is an odd number of letters $a$.

$L_3$ consists of all words which start with an \underline{odd} number of
letters $a$ and after first letter $b$ (if there is one) there
is an odd number of letters $a$.

It is easy to see that $L_1=L_2\bigcup L_3$.

The minimal automatons $G_2$ and $G_3$ are shown in
Fig.\ref{Bilde3} and Fig.\ref{Bilde4}.
They do not contain any of the ``forbidden constructions''
of Theorem \ref{T14}.
Therefore, $L_2$ and $L_3$ can be recognized by a QFA
and we get

\begin{theorem}
There are two languages $L_2$ and $L_3$ which are recognizable
by a QFA but the 
union of them $L_1=L_2\bigcup L_3$ is not recognizable by a QFA.
\end{theorem}

\begin{corollary}
\label{cor2}
The class of languages recognizable by a QFA is not closed under union.
\end{corollary}

This answers a question of Brodsky and Pippenger \cite{BP 99}.

As $L_2\bigcap L_3=\emptyset$ then also $L_1=L_2 \Delta L_3$.
So the class of languages recognizable by QFA
is not closed under symmetric difference.
From this and from the fact that this class
is closed under complement, it easily follows:

\begin{corollary}
\label{cor3}
The class of languages recognizable by a QFA is not closed under any binary
boolean operation where both arguments are significant.
\end{corollary}

\subsection{Another construction of QFAs}

Instead of using the general construction of Theorem \ref{T14}, 
we can also use a construction specific to languages $L_2$ and $L_3$.
This gives simpler QFAs and achieves
a better probability of correct answer.
(Theorem \ref{T14} gives QFAs for $L_2$ and $L_3$ 
with the probability of correct answer 3/5. 
Our construction below achieves the probability of 
correct answer 2/3.)


\begin{figure}[htb]
  \begin{minipage}{0.4\textwidth}
    \centering
    \begin{Large}
\special{em:linewidth 0.4pt}
\unitlength 1mm
\linethickness{0.4pt}
\begin{picture}(70.00,47.00)
\emline{50.00}{47.00}{1}{51.59}{46.82}{2}
\emline{51.59}{46.82}{3}{53.09}{46.28}{4}
\emline{53.09}{46.28}{5}{54.43}{45.42}{6}
\emline{54.43}{45.42}{7}{55.54}{44.27}{8}
\emline{55.54}{44.27}{9}{56.37}{42.91}{10}
\emline{56.37}{42.91}{11}{56.86}{41.39}{12}
\emline{56.86}{41.39}{13}{57.00}{39.80}{14}
\emline{57.00}{39.80}{15}{56.77}{38.22}{16}
\emline{56.77}{38.22}{17}{56.19}{36.73}{18}
\emline{56.19}{36.73}{19}{55.29}{35.42}{20}
\emline{55.29}{35.42}{21}{54.11}{34.34}{22}
\emline{54.11}{34.34}{23}{52.72}{33.55}{24}
\emline{52.72}{33.55}{25}{51.19}{33.10}{26}
\emline{51.19}{33.10}{27}{49.60}{33.01}{28}
\emline{49.60}{33.01}{29}{48.03}{33.28}{30}
\emline{48.03}{33.28}{31}{46.56}{33.90}{32}
\emline{46.56}{33.90}{33}{45.27}{34.84}{34}
\emline{45.27}{34.84}{35}{44.22}{36.05}{36}
\emline{44.22}{36.05}{37}{43.48}{37.46}{38}
\emline{43.48}{37.46}{39}{43.07}{39.00}{40}
\emline{43.07}{39.00}{41}{43.03}{40.60}{42}
\emline{43.03}{40.60}{43}{43.34}{42.16}{44}
\emline{43.34}{42.16}{45}{44.01}{43.61}{46}
\emline{44.01}{43.61}{47}{44.98}{44.88}{48}
\emline{44.98}{44.88}{49}{46.22}{45.89}{50}
\emline{46.22}{45.89}{51}{47.65}{46.59}{52}
\emline{47.65}{46.59}{53}{50.00}{47.00}{54}
\put(50.00,40.00){\circle{10.00}}
\put(20.00,40.00){\circle{10.00}}
\put(43.67,43.00){\vector(4,-1){0.2}}
\emline{24.00}{43.00}{55}{26.42}{43.58}{56}
\emline{26.42}{43.58}{57}{28.85}{44.00}{58}
\emline{28.85}{44.00}{59}{31.30}{44.25}{60}
\emline{31.30}{44.25}{61}{36.21}{44.25}{62}
\emline{36.21}{44.25}{63}{38.69}{44.00}{64}
\emline{38.69}{44.00}{65}{41.17}{43.58}{66}
\emline{41.17}{43.58}{67}{43.67}{43.00}{68}
\put(24.00,37.00){\vector(-4,1){0.2}}
\emline{43.67}{37.00}{69}{41.03}{36.42}{70}
\emline{41.03}{36.42}{71}{38.44}{36.00}{72}
\emline{38.44}{36.00}{73}{35.90}{35.75}{74}
\emline{35.90}{35.75}{75}{30.98}{35.75}{76}
\emline{30.98}{35.75}{77}{28.60}{36.00}{78}
\emline{28.60}{36.00}{79}{26.28}{36.42}{80}
\emline{26.28}{36.42}{81}{24.00}{37.00}{82}
\put(50.00,40.00){\makebox(0,0)[cc]{$q_1$}}
\put(20.00,40.00){\makebox(0,0)[cc]{$q_4$}}
\emline{20.00}{17.00}{83}{21.59}{16.82}{84}
\emline{21.59}{16.82}{85}{23.09}{16.28}{86}
\emline{23.09}{16.28}{87}{24.43}{15.42}{88}
\emline{24.43}{15.42}{89}{25.54}{14.27}{90}
\emline{25.54}{14.27}{91}{26.37}{12.91}{92}
\emline{26.37}{12.91}{93}{26.86}{11.39}{94}
\emline{26.86}{11.39}{95}{27.00}{9.80}{96}
\emline{27.00}{9.80}{97}{26.77}{8.22}{98}
\emline{26.77}{8.22}{99}{26.19}{6.73}{100}
\emline{26.19}{6.73}{101}{25.29}{5.42}{102}
\emline{25.29}{5.42}{103}{24.11}{4.34}{104}
\emline{24.11}{4.34}{105}{22.72}{3.55}{106}
\emline{22.72}{3.55}{107}{21.19}{3.10}{108}
\emline{21.19}{3.10}{109}{19.60}{3.01}{110}
\emline{19.60}{3.01}{111}{18.03}{3.28}{112}
\emline{18.03}{3.28}{113}{16.56}{3.90}{114}
\emline{16.56}{3.90}{115}{15.27}{4.84}{116}
\emline{15.27}{4.84}{117}{14.22}{6.05}{118}
\emline{14.22}{6.05}{119}{13.48}{7.46}{120}
\emline{13.48}{7.46}{121}{13.07}{9.00}{122}
\emline{13.07}{9.00}{123}{13.03}{10.60}{124}
\emline{13.03}{10.60}{125}{13.34}{12.16}{126}
\emline{13.34}{12.16}{127}{14.01}{13.61}{128}
\emline{14.01}{13.61}{129}{14.98}{14.88}{130}
\emline{14.98}{14.88}{131}{16.22}{15.89}{132}
\emline{16.22}{15.89}{133}{17.65}{16.59}{134}
\emline{17.65}{16.59}{135}{20.00}{17.00}{136}
\put(20.00,10.00){\circle{10.00}}
\put(50.00,10.00){\circle{10.00}}
\put(26.33,7.00){\vector(-4,1){0.2}}
\emline{46.00}{7.00}{137}{43.58}{6.42}{138}
\emline{43.58}{6.42}{139}{41.15}{6.00}{140}
\emline{41.15}{6.00}{141}{38.70}{5.75}{142}
\emline{38.70}{5.75}{143}{33.79}{5.75}{144}
\emline{33.79}{5.75}{145}{31.31}{6.00}{146}
\emline{31.31}{6.00}{147}{28.83}{6.42}{148}
\emline{28.83}{6.42}{149}{26.33}{7.00}{150}
\put(46.00,13.00){\vector(4,-1){0.2}}
\emline{26.33}{13.00}{151}{28.97}{13.58}{152}
\emline{28.97}{13.58}{153}{31.56}{14.00}{154}
\emline{31.56}{14.00}{155}{34.10}{14.25}{156}
\emline{34.10}{14.25}{157}{39.02}{14.25}{158}
\emline{39.02}{14.25}{159}{41.40}{14.00}{160}
\emline{41.40}{14.00}{161}{43.72}{13.58}{162}
\emline{43.72}{13.58}{163}{46.00}{13.00}{164}
\put(20.00,10.00){\makebox(0,0)[cc]{$q_3$}}
\put(50.00,10.00){\makebox(0,0)[cc]{$q_2$}}
\put(54.00,7.00){\vector(-4,-1){0.2}}
\emline{54.00}{13.00}{165}{56.24}{12.55}{166}
\emline{56.24}{12.55}{167}{58.11}{12.09}{168}
\emline{58.11}{12.09}{169}{59.62}{11.64}{170}
\emline{59.62}{11.64}{171}{60.76}{11.18}{172}
\emline{60.76}{11.18}{173}{61.53}{10.73}{174}
\emline{61.53}{10.73}{175}{61.93}{10.27}{176}
\emline{61.93}{10.27}{177}{61.97}{9.82}{178}
\emline{61.97}{9.82}{179}{61.64}{9.36}{180}
\emline{61.64}{9.36}{181}{60.94}{8.91}{182}
\emline{60.94}{8.91}{183}{59.88}{8.45}{184}
\emline{59.88}{8.45}{185}{58.44}{8.00}{186}
\emline{58.44}{8.00}{187}{56.64}{7.55}{188}
\emline{56.64}{7.55}{189}{54.00}{7.00}{190}
\put(13.67,7.00){\vector(4,-1){0.2}}
\emline{13.67}{13.00}{191}{11.46}{12.48}{192}
\emline{11.46}{12.48}{193}{9.67}{11.97}{194}
\emline{9.67}{11.97}{195}{8.30}{11.45}{196}
\emline{8.30}{11.45}{197}{7.34}{10.93}{198}
\emline{7.34}{10.93}{199}{6.80}{10.41}{200}
\emline{6.80}{10.41}{201}{6.67}{9.90}{202}
\emline{6.67}{9.90}{203}{6.97}{9.38}{204}
\emline{6.97}{9.38}{205}{7.67}{8.86}{206}
\emline{7.67}{8.86}{207}{8.80}{8.34}{208}
\emline{8.80}{8.34}{209}{10.34}{7.83}{210}
\emline{10.34}{7.83}{211}{13.67}{7.00}{212}
\put(5.00,25.00){\circle{10.00}}
\put(5.00,25.00){\makebox(0,0)[cc]{$q_5$}}
\put(9.00,22.00){\vector(-4,-1){0.2}}
\emline{9.00}{28.00}{213}{11.24}{27.55}{214}
\emline{11.24}{27.55}{215}{13.11}{27.09}{216}
\emline{13.11}{27.09}{217}{14.62}{26.64}{218}
\emline{14.62}{26.64}{219}{15.76}{26.18}{220}
\emline{15.76}{26.18}{221}{16.53}{25.73}{222}
\emline{16.53}{25.73}{223}{16.93}{25.27}{224}
\emline{16.93}{25.27}{225}{16.97}{24.82}{226}
\emline{16.97}{24.82}{227}{16.64}{24.36}{228}
\emline{16.64}{24.36}{229}{15.94}{23.91}{230}
\emline{15.94}{23.91}{231}{14.88}{23.45}{232}
\emline{14.88}{23.45}{233}{13.44}{23.00}{234}
\emline{13.44}{23.00}{235}{11.64}{22.55}{236}
\emline{11.64}{22.55}{237}{9.00}{22.00}{238}
\put(12.67,30.00){\makebox(0,0)[cc]{a,b}}
\put(53.00,14.00){\vector(-1,-3){0.2}}
\emline{53.00}{33.67}{239}{53.84}{31.36}{240}
\emline{53.84}{31.36}{241}{54.45}{29.04}{242}
\emline{54.45}{29.04}{243}{54.84}{26.72}{244}
\emline{54.84}{26.72}{245}{55.00}{24.38}{246}
\emline{55.00}{24.38}{247}{54.93}{22.04}{248}
\emline{54.93}{22.04}{249}{54.63}{19.69}{250}
\emline{54.63}{19.69}{251}{54.11}{17.32}{252}
\emline{54.11}{17.32}{253}{53.00}{14.00}{254}
\put(6.00,40.00){\makebox(0,0)[cc]{$b$}}
\put(35.00,47.00){\makebox(0,0)[cc]{$a$}}
\put(35.00,39.00){\makebox(0,0)[cc]{$a$}}
\put(57.00,25.00){\makebox(0,0)[cc]{$b$}}
\put(58.00,15.00){\makebox(0,0)[cc]{$b$}}
\put(35.00,17.00){\makebox(0,0)[cc]{$a$}}
\put(35.00,8.00){\makebox(0,0)[cc]{$a$}}
\put(10.00,14.67){\makebox(0,0)[cc]{$b$}}
\put(16.00,43.00){\vector(1,0){0.2}}
\emline{6.00}{43.00}{255}{16.00}{43.00}{256}
\put(5.00,30.00){\vector(-1,-4){0.2}}
\emline{15.00}{40.00}{257}{12.65}{39.57}{258}
\emline{12.65}{39.57}{259}{10.62}{38.83}{260}
\emline{10.62}{38.83}{261}{8.89}{37.78}{262}
\emline{8.89}{37.78}{263}{7.47}{36.42}{264}
\emline{7.47}{36.42}{265}{6.36}{34.75}{266}
\emline{6.36}{34.75}{267}{5.56}{32.78}{268}
\emline{5.56}{32.78}{269}{5.00}{30.00}{270}
\end{picture}
    \end{Large}
    \setlength{\abovecaptionskip}{0pt}
    \caption{Automaton $G_2$}
    \label{Bilde3}
  \end{minipage}
  \hspace{0.1\textwidth}
  \begin{minipage}{0.4\textwidth}
    \centering
    \begin{Large}
\special{em:linewidth 0.4pt}
\unitlength 1mm
\linethickness{0.4pt}
\begin{picture}(70.00,47.00)
\emline{50.00}{47.00}{1}{51.59}{46.82}{2}
\emline{51.59}{46.82}{3}{53.09}{46.28}{4}
\emline{53.09}{46.28}{5}{54.43}{45.42}{6}
\emline{54.43}{45.42}{7}{55.54}{44.27}{8}
\emline{55.54}{44.27}{9}{56.37}{42.91}{10}
\emline{56.37}{42.91}{11}{56.86}{41.39}{12}
\emline{56.86}{41.39}{13}{57.00}{39.80}{14}
\emline{57.00}{39.80}{15}{56.77}{38.22}{16}
\emline{56.77}{38.22}{17}{56.19}{36.73}{18}
\emline{56.19}{36.73}{19}{55.29}{35.42}{20}
\emline{55.29}{35.42}{21}{54.11}{34.34}{22}
\emline{54.11}{34.34}{23}{52.72}{33.55}{24}
\emline{52.72}{33.55}{25}{51.19}{33.10}{26}
\emline{51.19}{33.10}{27}{49.60}{33.01}{28}
\emline{49.60}{33.01}{29}{48.03}{33.28}{30}
\emline{48.03}{33.28}{31}{46.56}{33.90}{32}
\emline{46.56}{33.90}{33}{45.27}{34.84}{34}
\emline{45.27}{34.84}{35}{44.22}{36.05}{36}
\emline{44.22}{36.05}{37}{43.48}{37.46}{38}
\emline{43.48}{37.46}{39}{43.07}{39.00}{40}
\emline{43.07}{39.00}{41}{43.03}{40.60}{42}
\emline{43.03}{40.60}{43}{43.34}{42.16}{44}
\emline{43.34}{42.16}{45}{44.01}{43.61}{46}
\emline{44.01}{43.61}{47}{44.98}{44.88}{48}
\emline{44.98}{44.88}{49}{46.22}{45.89}{50}
\emline{46.22}{45.89}{51}{47.65}{46.59}{52}
\emline{47.65}{46.59}{53}{50.00}{47.00}{54}
\put(50.00,40.00){\circle{10.00}}
\put(20.00,40.00){\circle{10.00}}
\put(43.67,43.00){\vector(4,-1){0.2}}
\emline{24.00}{43.00}{55}{26.42}{43.58}{56}
\emline{26.42}{43.58}{57}{28.85}{44.00}{58}
\emline{28.85}{44.00}{59}{31.30}{44.25}{60}
\emline{31.30}{44.25}{61}{36.21}{44.25}{62}
\emline{36.21}{44.25}{63}{38.69}{44.00}{64}
\emline{38.69}{44.00}{65}{41.17}{43.58}{66}
\emline{41.17}{43.58}{67}{43.67}{43.00}{68}
\put(24.00,37.00){\vector(-4,1){0.2}}
\emline{43.67}{37.00}{69}{41.03}{36.42}{70}
\emline{41.03}{36.42}{71}{38.44}{36.00}{72}
\emline{38.44}{36.00}{73}{35.90}{35.75}{74}
\emline{35.90}{35.75}{75}{30.98}{35.75}{76}
\emline{30.98}{35.75}{77}{28.60}{36.00}{78}
\emline{28.60}{36.00}{79}{26.28}{36.42}{80}
\emline{26.28}{36.42}{81}{24.00}{37.00}{82}
\put(50.00,40.00){\makebox(0,0)[cc]{$q_1$}}
\put(20.00,40.00){\makebox(0,0)[cc]{$q_4$}}
\emline{20.00}{17.00}{83}{21.59}{16.82}{84}
\emline{21.59}{16.82}{85}{23.09}{16.28}{86}
\emline{23.09}{16.28}{87}{24.43}{15.42}{88}
\emline{24.43}{15.42}{89}{25.54}{14.27}{90}
\emline{25.54}{14.27}{91}{26.37}{12.91}{92}
\emline{26.37}{12.91}{93}{26.86}{11.39}{94}
\emline{26.86}{11.39}{95}{27.00}{9.80}{96}
\emline{27.00}{9.80}{97}{26.77}{8.22}{98}
\emline{26.77}{8.22}{99}{26.19}{6.73}{100}
\emline{26.19}{6.73}{101}{25.29}{5.42}{102}
\emline{25.29}{5.42}{103}{24.11}{4.34}{104}
\emline{24.11}{4.34}{105}{22.72}{3.55}{106}
\emline{22.72}{3.55}{107}{21.19}{3.10}{108}
\emline{21.19}{3.10}{109}{19.60}{3.01}{110}
\emline{19.60}{3.01}{111}{18.03}{3.28}{112}
\emline{18.03}{3.28}{113}{16.56}{3.90}{114}
\emline{16.56}{3.90}{115}{15.27}{4.84}{116}
\emline{15.27}{4.84}{117}{14.22}{6.05}{118}
\emline{14.22}{6.05}{119}{13.48}{7.46}{120}
\emline{13.48}{7.46}{121}{13.07}{9.00}{122}
\emline{13.07}{9.00}{123}{13.03}{10.60}{124}
\emline{13.03}{10.60}{125}{13.34}{12.16}{126}
\emline{13.34}{12.16}{127}{14.01}{13.61}{128}
\emline{14.01}{13.61}{129}{14.98}{14.88}{130}
\emline{14.98}{14.88}{131}{16.22}{15.89}{132}
\emline{16.22}{15.89}{133}{17.65}{16.59}{134}
\emline{17.65}{16.59}{135}{20.00}{17.00}{136}
\put(20.00,10.00){\circle{10.00}}
\put(50.00,10.00){\circle{10.00}}
\put(26.33,7.00){\vector(-4,1){0.2}}
\emline{46.00}{7.00}{137}{43.58}{6.42}{138}
\emline{43.58}{6.42}{139}{41.15}{6.00}{140}
\emline{41.15}{6.00}{141}{38.70}{5.75}{142}
\emline{38.70}{5.75}{143}{33.79}{5.75}{144}
\emline{33.79}{5.75}{145}{31.31}{6.00}{146}
\emline{31.31}{6.00}{147}{28.83}{6.42}{148}
\emline{28.83}{6.42}{149}{26.33}{7.00}{150}
\put(46.00,13.00){\vector(4,-1){0.2}}
\emline{26.33}{13.00}{151}{28.97}{13.58}{152}
\emline{28.97}{13.58}{153}{31.56}{14.00}{154}
\emline{31.56}{14.00}{155}{34.10}{14.25}{156}
\emline{34.10}{14.25}{157}{39.02}{14.25}{158}
\emline{39.02}{14.25}{159}{41.40}{14.00}{160}
\emline{41.40}{14.00}{161}{43.72}{13.58}{162}
\emline{43.72}{13.58}{163}{46.00}{13.00}{164}
\put(20.00,10.00){\makebox(0,0)[cc]{$q_3$}}
\put(50.00,10.00){\makebox(0,0)[cc]{$q_2$}}
\put(54.00,7.00){\vector(-4,-1){0.2}}
\emline{54.00}{13.00}{165}{56.24}{12.55}{166}
\emline{56.24}{12.55}{167}{58.11}{12.09}{168}
\emline{58.11}{12.09}{169}{59.62}{11.64}{170}
\emline{59.62}{11.64}{171}{60.76}{11.18}{172}
\emline{60.76}{11.18}{173}{61.53}{10.73}{174}
\emline{61.53}{10.73}{175}{61.93}{10.27}{176}
\emline{61.93}{10.27}{177}{61.97}{9.82}{178}
\emline{61.97}{9.82}{179}{61.64}{9.36}{180}
\emline{61.64}{9.36}{181}{60.94}{8.91}{182}
\emline{60.94}{8.91}{183}{59.88}{8.45}{184}
\emline{59.88}{8.45}{185}{58.44}{8.00}{186}
\emline{58.44}{8.00}{187}{56.64}{7.55}{188}
\emline{56.64}{7.55}{189}{54.00}{7.00}{190}
\put(13.67,7.00){\vector(4,-1){0.2}}
\emline{13.67}{13.00}{191}{11.46}{12.48}{192}
\emline{11.46}{12.48}{193}{9.67}{11.97}{194}
\emline{9.67}{11.97}{195}{8.30}{11.45}{196}
\emline{8.30}{11.45}{197}{7.34}{10.93}{198}
\emline{7.34}{10.93}{199}{6.80}{10.41}{200}
\emline{6.80}{10.41}{201}{6.67}{9.90}{202}
\emline{6.67}{9.90}{203}{6.97}{9.38}{204}
\emline{6.97}{9.38}{205}{7.67}{8.86}{206}
\emline{7.67}{8.86}{207}{8.80}{8.34}{208}
\emline{8.80}{8.34}{209}{10.34}{7.83}{210}
\emline{10.34}{7.83}{211}{13.67}{7.00}{212}
\put(5.00,25.00){\circle{10.00}}
\put(5.00,25.00){\makebox(0,0)[cc]{$q_5$}}
\put(9.00,22.00){\vector(-4,-1){0.2}}
\emline{9.00}{28.00}{213}{11.24}{27.55}{214}
\emline{11.24}{27.55}{215}{13.11}{27.09}{216}
\emline{13.11}{27.09}{217}{14.62}{26.64}{218}
\emline{14.62}{26.64}{219}{15.76}{26.18}{220}
\emline{15.76}{26.18}{221}{16.53}{25.73}{222}
\emline{16.53}{25.73}{223}{16.93}{25.27}{224}
\emline{16.93}{25.27}{225}{16.97}{24.82}{226}
\emline{16.97}{24.82}{227}{16.64}{24.36}{228}
\emline{16.64}{24.36}{229}{15.94}{23.91}{230}
\emline{15.94}{23.91}{231}{14.88}{23.45}{232}
\emline{14.88}{23.45}{233}{13.44}{23.00}{234}
\emline{13.44}{23.00}{235}{11.64}{22.55}{236}
\emline{11.64}{22.55}{237}{9.00}{22.00}{238}
\put(12.67,30.00){\makebox(0,0)[cc]{a,b}}
\put(53.00,14.00){\vector(-1,-3){0.2}}
\emline{53.00}{33.67}{239}{53.84}{31.36}{240}
\emline{53.84}{31.36}{241}{54.45}{29.04}{242}
\emline{54.45}{29.04}{243}{54.84}{26.72}{244}
\emline{54.84}{26.72}{245}{55.00}{24.38}{246}
\emline{55.00}{24.38}{247}{54.93}{22.04}{248}
\emline{54.93}{22.04}{249}{54.63}{19.69}{250}
\emline{54.63}{19.69}{251}{54.11}{17.32}{252}
\emline{54.11}{17.32}{253}{53.00}{14.00}{254}
\put(6.00,40.00){\makebox(0,0)[cc]{$b$}}
\put(35.00,47.00){\makebox(0,0)[cc]{$a$}}
\put(35.00,39.00){\makebox(0,0)[cc]{$a$}}
\put(57.00,25.00){\makebox(0,0)[cc]{$b$}}
\put(58.00,15.00){\makebox(0,0)[cc]{$b$}}
\put(35.00,17.00){\makebox(0,0)[cc]{$a$}}
\put(35.00,8.00){\makebox(0,0)[cc]{$a$}}
\put(10.00,14.67){\makebox(0,0)[cc]{$b$}}
\put(5.00,30.00){\vector(-1,-4){0.2}}
\emline{15.00}{40.00}{255}{12.65}{39.57}{256}
\emline{12.65}{39.57}{257}{10.62}{38.83}{258}
\emline{10.62}{38.83}{259}{8.89}{37.78}{260}
\emline{8.89}{37.78}{261}{7.47}{36.42}{262}
\emline{7.47}{36.42}{263}{6.36}{34.75}{264}
\emline{6.36}{34.75}{265}{5.56}{32.78}{266}
\emline{5.56}{32.78}{267}{5.00}{30.00}{268}
\put(56.33,43.00){\vector(-1,0){0.2}}
\emline{66.33}{43.00}{269}{56.33}{43.00}{270}
\end{picture}
    \end{Large}
    \setlength{\abovecaptionskip}{0pt}
    \caption{Automaton $G_3$}
    \label{Bilde4}
  \end{minipage}
\end{figure}

We construct two quantum automata $K_2$ and $K_3$ which
recognize languages $L_2$ and $L_3$. Like $G_2$ and $G_3$
they differ only in a starting state.

The automaton $K_2$ consists of 8 states:
$q_1$, $q_2$, $q_3$, $q_4$, $q_5$, $q_6$, $q_7$, $q_8$, where
$Q_{non}=\{q_1, q_2, q_3, q_4\}$, $Q_{acc}=\{q_5, q_8\}$, 
$Q_{rej}=\{q_6, q_7\}$.
The unitary transform matrices $V_\kappa$, $V_a$, $V_b$ and $V_\$$ are:
$$V_\kappa=
\left (
\begin{array}{cccccccc}
\sqrt\frac{2}{3}&\sqrt\frac{1}{3}&0&0&0&0&0&0\\
\sqrt\frac{1}{3}&-\sqrt\frac{2}{3}&0&0&0&0&0&0\\
0&0&-\sqrt\frac{2}{3}&\sqrt\frac{1}{3}&0&0&0&0\\
0&0&\sqrt\frac{2}{3}&\sqrt\frac{1}{3}&0&0&0&0\\
0&0&0&0&1&0&0&0\\
0&0&0&0&0&1&0&0\\
0&0&0&0&0&0&1&0\\
0&0&0&0&0&0&0&1
\end{array}
\right ),
V_a=
\left (
\begin{array}{cccccccc}
0&0&0&1&0&0&0&0\\
0&0&1&0&0&0&0&0\\
0&1&0&0&0&0&0&0\\
1&0&0&0&0&0&0&0\\
0&0&0&0&1&0&0&0\\
0&0&0&0&0&1&0&0\\
0&0&0&0&0&0&1&0\\
0&0&0&0&0&0&0&1
\end{array}
\right )
$$

$$V_b=
\left (
\begin{array}{cccccccc}
0&0&0&0&\sqrt\frac{1}{2}&\sqrt\frac{1}{2}&0&0\\
0&1&0&0&0&0&0&0\\
0&0&1&0&0&0&0&0\\
0&0&0&0&0&0&1&0\\
\sqrt\frac{1}{2}&0&0&0&\frac{1}{2}&-\frac{1}{2}&0&0\\
\sqrt\frac{1}{2}&0&0&0&-\frac{1}{2}&\frac{1}{2}&0&0\\
0&0&0&1&0&0&0&0\\
0&0&0&0&0&0&0&1
\end{array}
\right ),
V_\$=
\left (
\begin{array}{cccccccc}
0&0&0&0&1&0&0&0\\
0&0&0&0&0&1&0&0\\
0&0&0&0&0&0&0&1\\
0&0&0&0&0&0&1&0\\
1&0&0&0&0&0&0&0\\
0&1&0&0&0&0&0&0\\
0&0&0&1&0&0&0&0\\
0&0&1&0&0&0&0&0
\end{array}
\right )
$$

The starting state is $q_1$ for $K_2$
and $q_4$ for $K_3$.
Next, we show that $K_2$ works similarly to $G_2$. 

The state $q_1$ in $G_2$ corresponds to 
$\psi_1=\sqrt\frac{2}{3}\ket{q_1}+\sqrt\frac{1}{3}\ket{q_2}$ in $K_2$.

The state $q_2$ in $G_2$ corresponds to 
$\psi_2=\sqrt\frac{1}{3}\ket{q_2}$ in $K_2$.

The state $q_3$ in $G_2$ corresponds to 
$\psi_3=\sqrt\frac{1}{3}\ket{q_3}$ in $K_2$.

The state $q_4$ in $G_2$ corresponds to 
$\psi_4=\sqrt\frac{2}{3}\ket{q_4}+\sqrt\frac{1}{3}\ket{q_3}$ in $K_2$.

%
%
%
\begin{enumerate}
\item
After reading the left endmarker $\kappa$ $K_2$ is in the state 
$V_\kappa'(\ket{q_1})=\psi_1$. 
$G_2$ is in its starting state $q_1$.
\item
If by reading the letter $a$ the automaton $G_2$ passes from $q_1$ to $q_4$
or back then the state of $K_2$ changes from $\psi_1$ to $\psi_4$
or back.
\item
If $K_2$ receives the letter $b$ in the state $\psi_4$
then it rejects the input with probability $\frac{2}{3}$.
This is correct because $G_2$ passes from $q_4$ to 
the "all rejecting" state $q_5$.
\item
If $G_2$ receives the letter $b$ in the state $q_1$
it passes to $q_3$. 
If $K_2$ receives the letter $b$ in the state $\psi_1$, 
it passes to the state
$\frac{1}{\sqrt{3}}\ket{q_2}+\frac{1}{\sqrt{3}}\ket{q_5}+\frac{1}{\sqrt{3}}\ket{q_6}$
and after the measurement accepts the input with
probability $\frac{1}{3}$,
rejects the input with the same probability $\frac{1}{3}$,
or continues in the state $\psi_2$.
\item
If by reading the letter $a$ the automaton $G_2$ passes from $q_2$ to $q_3$
or back then the state of $K_2$ changes from $\psi_2$ to $\psi_3$
or back. 
By reading the letter $b$ $G_2$ passes from
$q_2$ to $q_2$ and from $q_3$ to $q_3$.
$K_2$ passes from $\psi_2$ to $\psi_2$ and from $\psi_3$ to $\psi_3$.
\item
If $K_2$ receives the right endmarker in state $\psi_1$
then the input is accepted with probability $\frac{2}{3}$.
\item
If $K_2$ receives the right endmarker in state $\psi_2$
then the input is rejected with probability $\frac{1}{3}$
and as it was rejected with the same probability so far,
the total probability to reject the input is $\frac{2}{3}$.
\item
If $K_2$ receives the right endmarker in state $\psi_3$
then the input is accepted with probability $\frac{1}{3}$
and as it was accepted with the same probability so far
the total probability to accept the input is $\frac{2}{3}$.
\item
If $K_2$ receives the right endmarker in state $\psi_4$
then the input is rejected with probability $\frac{2}{3}$.
\end{enumerate}

This shows that, whenever $G_2$ is in a state $q_i$ 
($i\in\{1, 2, 3, 4\}$), $K_2$ is in the corresponding state $\psi_i$.
Also, this shows that $K_2$ accepts input with probability
$\frac{2}{3}$ iff it receives right endmarker $\$$ in one of
states $\psi_1$ or $\psi_3$ corresponding to
$q_1$ and $q_3$, the only accepting states in $G_1$.
So, we can conclude
that $K_2$ accepts the language $L_2$ with
probability $\frac{2}{3}$.
Similarly, we can show that $K_3$ accepts $L_3$
with probability $\frac{2}{3}$.


Thus, we have shown

\begin{theorem}
\label{cor1}
There are two languages $L_2$ and $L_3$ which are recognizable
by a QFA with probability $\frac{2}{3}$ but
the union of them $L_1=L_2\bigcup L_3$ is not recognizable with a QFA
(with any probability $1/2+\epsilon$, $\epsilon>0$).
\end{theorem}

\subsection{On accepting probabilities}

The probabilities for $L_2$ and $L_3$ achieved in Theorem \ref{cor1} are 
the best possible, as shown by the following theorem.

\begin{theorem}
\label{T31}
If 2 languages $L_1$ and $L_2$ are recognizable by a QFA with
probabilities $p_1$ and $p_2$ and $\frac{1}{p_1}+\frac{1}{p_2}<3$
then $L=L_1\bigcup L_2$ is also recognizable by QFA with
probability $\frac{2p_1p_2}{p_1+p_2+p_1p_2}$.
\end{theorem}

\begin{proof}
We have a QFA $K_1$ which accepts $L_1$ with probability $p_1$ and
a QFA $K_2$ which accepts $L_2$ with probability $p_2$.
We will make a QFA $K$ which will work like this:
\begin{enumerate}
\item
Runs $K_1$ with probability $\frac{p_2}{p_1+p_2+p_1p_2}$,
\item
Runs $K_2$ with probability $\frac{p_1}{p_1+p_2+p_1p_2}$,
\item
Accepts input with probability $\frac{p_1p_2}{p_1+p_2+p_1p_2}$.
\end{enumerate}


\begin{enumerate}
\item
$w\cin L_1$ and $w\cin L_2\longrightarrow$ input is accepted
with probability

$$\frac{p_2}{p_1+p_2+p_1p_2}\cdot p_1+\frac{p_1}{p_1+p_2+p_1p_2}\cdot p_2+
\frac{p_1p_2}{p_1+p_2+p_1p_2}\cdot 1=\frac{3p_1p_2}{p_1+p_2+p_1p_2}$$
\item
$w\cin L_1$ and $w\cnotin L_2\longrightarrow$ input is accepted
with probability at least

$$\frac{p_2}{p_1+p_2+p_1p_2}\cdot p_1+\frac{p_1p_2}{p_1+p_2+p_1p_2}\cdot 1=
\frac{2p_1p_2}{p_1+p_2+p_1p_2}$$
\item
$w\cnotin L_1$ and $w\cin L_2\longrightarrow$ input is accepted
with probability at least

$$\frac{p_1}{p_1+p_2+p_1p_2}\cdot p_2+\frac{p_1p_2}{p_1+p_2+p_1p_2}\cdot 1=
\frac{2p_1p_2}{p_1+p_2+p_1p_2}$$
\item
$w\cnotin L_1$ and $w\cnotin L_2\longrightarrow$ input is rejected
with probability at least

$$\frac{p_2}{p_1+p_2+p_1p_2}\cdot p_1+\frac{p_1}{p_1+p_2+p_1p_2}\cdot p_2=
\frac{2p_1p_2}{p_1+p_2+p_1p_2}$$
\end{enumerate}
So automaton $K$ recognizes $L$ with the probability
at least
$$\frac{2p_1p_2}{p_1+p_2+p_1p_2}=
\frac{1}{2}+\frac{3-(\frac{1}{p_1}+
\frac{1}{p_2})}{4(1+\frac{1}{p_1}+\frac{1}{p_2})}>\frac{1}{2}$$
\qed
\end{proof}

All this has also a nice geometric interpretation.
We are going to build a linear
function $f$ from probabilities $x_1$, $x_2$
to probability $x$ such that
$f(p_1,p_2)\geq\frac{1}{2}+\varepsilon$,
$f(p_1,0)\geq\frac{1}{2}+\varepsilon$, 
$f(0,p_2)\geq\frac{1}{2}+\varepsilon$, 
$f(1-p_1,1-p_2)\leq\frac{1}{2}-\varepsilon$.
Geometrically we consider a plane $x,y$ where each word $w$ is
located in a point $(x,y)$, where $x$ is probability
that $K_1$ accepts $w$ and $y$ is the probability,
that $K_2$ accepts $w$.

$S_1$ is the place where lies all words that do not belong to $L$.
$S_2$ is the place where lies all words that belong to $L$.

If we can (Fig.\ref{Bilde5}) separate these two parts with a line $ax+by=c$
then we can construct automaton "$K=aK_1+bK_2$"
with $c$ as isolated cut point. 
If we can not (Fig.\ref{Bilde6}) then this method doesn't help.
And as it was shown higher sometimes none of other methods can help, too.

Case when $p_1=p_2=\frac{2}{3}$ (Fig.\ref{Bilde9}) is the limit case.
If any of probabilities were a little bit greater
then this method would help.

Sometimes it may be that there are no words $w$ such that
$K_1$ or $K_2$ would reject with probability $1-t$ or greater.
Then (Fig.\ref{Bilde7}) you can see that now it is easier to make
such a line so the condition $\frac{1}{p_1}+\frac{1}{p_2}<3$
can be weakened
(the probabilities in Fig.\ref{Bilde7}
are the same as in Fig.\ref{Bilde6}).
In the limit case when rejecting probabilities are only
$p_1$ and $p_2$, $S_1$ is the point $(1-p_1,1-p_2)$ (Fig.\ref{Bilde8}).
So with any $p_1$ and $p_2$ you can separate $S_1$ from $S_2$
with a line, from what it follows you can always
construct $K=K_1\bigcup K_2$.

Now it is clear that the languages $L_2$ and $L_3$ defined in
section 3 cannot be recognized with probability greater than $\frac{2}{3}$
so the construction presented there is best possible.

\begin{corollary}
If 2 languages $L_1$ and $L_2$ are recognizable by a QFA with
probabilities $p_1$ and $p_2$ and $p_1>2/3$ and $p_2>2/3$,
then $L=L_1\bigcup L_2$ is recognizable by QFA with probability $p_3>1/2$.
\end{corollary}

\begin{proof}
Follows immediately from Theorem \ref{T31}.
\qed
\end{proof}

\section{More "forbidden" constructions}

If we allow the "two cycles in a row" construction, 
Theorem \ref{T14} is not longer true.
More and more complicated "forbidden fragments" that
imply non-recognizability by a QFA are possible.

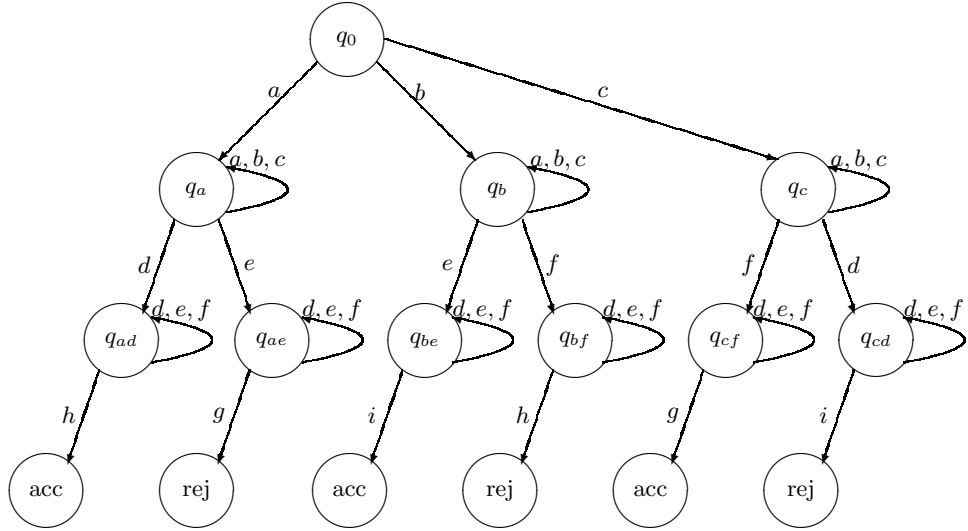
\begin{figure}[htb]
  \centering
\unitlength 1.00mm
\linethickness{0.4pt}
\begin{picture}(150.00,75.00)
\put(40.00,50.00){\circle{10.00}}
\put(40.00,50.00){\makebox(0,0)[cc]{$q_a$}}
\put(80.00,50.00){\circle{10.00}}
\put(80.00,50.00){\makebox(0,0)[cc]{$q_b$}}
\put(88.00,54.00){\makebox(0,0)[cc]{$a,b,c$}}
\put(84.00,53.00){\vector(-4,1){0.2}}
\bezier{132}(84.00,47.00)(100.00,50.00)(84.00,53.00)
\put(60.00,70.00){\circle{10.00}}
\put(69.67,63.00){\makebox(0,0)[cc]{$b$}}
\put(60.00,70.00){\makebox(0,0)[cc]{$q_0$}}
\put(30.00,30.00){\circle{10.00}}
\put(33.00,34.00){\vector(-1,-3){0.2}}
\multiput(37.00,46.00)(-0.12,-0.35){34}{\line(0,-1){0.35}}
\put(30.00,30.00){\makebox(0,0)[cc]{$q_{ad}$}}
\put(50.00,30.00){\circle{10.00}}
\put(47.00,34.00){\vector(1,-3){0.2}}
\multiput(43.00,46.00)(0.12,-0.35){34}{\line(0,-1){0.35}}
\put(50.00,30.00){\makebox(0,0)[cc]{$q_{ae}$}}
\put(33.00,40.00){\makebox(0,0)[cc]{$d$}}
\put(47.00,40.00){\makebox(0,0)[cc]{$e$}}
\put(70.33,30.00){\circle{10.00}}
\put(73.33,34.00){\vector(-1,-3){0.2}}
\multiput(77.33,46.00)(-0.12,-0.35){34}{\line(0,-1){0.35}}
\put(70.33,30.00){\makebox(0,0)[cc]{$q_{be}$}}
\put(90.33,30.00){\circle{10.00}}
\put(87.33,34.00){\vector(1,-3){0.2}}
\multiput(83.33,46.00)(0.12,-0.35){34}{\line(0,-1){0.35}}
\put(90.33,30.00){\makebox(0,0)[cc]{$q_{bf}$}}
\put(73.33,40.00){\makebox(0,0)[cc]{$e$}}
\put(87.33,40.00){\makebox(0,0)[cc]{$f$}}
\put(77.00,54.00){\vector(1,-1){0.2}}
\multiput(64.00,67.00)(0.12,-0.12){109}{\line(0,-1){0.12}}
\put(120.00,50.00){\circle{10.00}}
\put(120.00,50.00){\makebox(0,0)[cc]{$q_c$}}
\put(128.00,54.00){\makebox(0,0)[cc]{$a,b,c$}}
\put(124.00,53.00){\vector(-4,1){0.2}}
\bezier{132}(124.00,47.00)(140.00,50.00)(124.00,53.00)
\put(110.33,30.00){\circle{10.00}}
\put(113.33,34.00){\vector(-1,-3){0.2}}
\multiput(117.33,46.00)(-0.12,-0.35){34}{\line(0,-1){0.35}}
\put(110.33,30.00){\makebox(0,0)[cc]{$q_{cf}$}}
\put(130.33,30.00){\circle{10.00}}
\put(127.33,34.00){\vector(1,-3){0.2}}
\multiput(123.33,46.00)(0.12,-0.35){34}{\line(0,-1){0.35}}
\put(130.33,30.00){\makebox(0,0)[cc]{$q_{cd}$}}
\put(113.33,40.00){\makebox(0,0)[cc]{$f$}}
\put(127.33,40.00){\makebox(0,0)[cc]{$d$}}
\put(50.33,63.00){\makebox(0,0)[cc]{$a$}}
\put(43.00,54.00){\vector(-1,-1){0.2}}
\multiput(56.00,67.00)(-0.12,-0.12){109}{\line(0,-1){0.12}}
\put(48.00,54.00){\makebox(0,0)[cc]{$a,b,c$}}
\put(44.00,53.00){\vector(-4,1){0.2}}
\bezier{132}(44.00,47.00)(60.00,50.00)(44.00,53.00)
\put(94.00,33.00){\vector(-4,1){0.2}}
\bezier{132}(94.00,27.00)(110.00,30.00)(94.00,33.00)
\put(134.00,33.00){\vector(-4,1){0.2}}
\bezier{132}(134.00,27.00)(150.00,30.00)(134.00,33.00)
\put(54.00,33.00){\vector(-4,1){0.2}}
\bezier{132}(54.00,27.00)(70.00,30.00)(54.00,33.00)
\put(74.00,33.00){\vector(-4,1){0.2}}
\bezier{132}(74.00,27.00)(90.00,30.00)(74.00,33.00)
\put(114.00,33.00){\vector(-4,1){0.2}}
\bezier{132}(114.00,27.00)(130.00,30.00)(114.00,33.00)
\put(38.00,34.00){\makebox(0,0)[cc]{$d,e,f$}}
\put(34.00,33.00){\vector(-4,1){0.2}}
\bezier{132}(34.00,27.00)(50.00,30.00)(34.00,33.00)
\put(94.00,63.00){\makebox(0,0)[cc]{$c$}}
\put(58.00,34.00){\makebox(0,0)[cc]{$d,e,f$}}
\put(78.00,34.00){\makebox(0,0)[cc]{$d,e,f$}}
\put(98.00,34.00){\makebox(0,0)[cc]{$d,e,f$}}
\put(118.00,34.00){\makebox(0,0)[cc]{$d,e,f$}}
\put(138.00,34.00){\makebox(0,0)[cc]{$d,e,f$}}
\put(20.00,10.00){\circle{10.00}}
\put(23.00,14.00){\vector(-1,-3){0.2}}
\multiput(27.00,26.00)(-0.12,-0.35){34}{\line(0,-1){0.35}}
\put(20.00,10.00){\makebox(0,0)[cc]{acc}}
\put(40.00,10.00){\circle{10.00}}
\put(40.00,10.00){\makebox(0,0)[cc]{rej}}
\put(23.00,20.00){\makebox(0,0)[cc]{$h$}}
\put(60.33,10.00){\circle{10.00}}
\put(63.33,14.00){\vector(-1,-3){0.2}}
\multiput(67.33,26.00)(-0.12,-0.35){34}{\line(0,-1){0.35}}
\put(60.33,10.00){\makebox(0,0)[cc]{acc}}
\put(80.33,10.00){\circle{10.00}}
\put(80.33,10.00){\makebox(0,0)[cc]{rej}}
\put(63.33,20.00){\makebox(0,0)[cc]{$i$}}
\put(100.33,10.00){\circle{10.00}}
\put(103.33,14.00){\vector(-1,-3){0.2}}
\multiput(107.33,26.00)(-0.12,-0.35){34}{\line(0,-1){0.35}}
\put(100.33,10.00){\makebox(0,0)[cc]{acc}}
\put(120.33,10.00){\circle{10.00}}
\put(120.33,10.00){\makebox(0,0)[cc]{rej}}
\put(103.33,20.00){\makebox(0,0)[cc]{$g$}}
\put(43.00,14.00){\vector(-1,-3){0.2}}
\multiput(47.00,26.00)(-0.12,-0.35){34}{\line(0,-1){0.35}}
\put(43.00,20.00){\makebox(0,0)[cc]{$g$}}
\put(83.33,14.00){\vector(-1,-3){0.2}}
\multiput(87.33,26.00)(-0.12,-0.35){34}{\line(0,-1){0.35}}
\put(83.33,20.00){\makebox(0,0)[cc]{$h$}}
\put(123.33,14.00){\vector(-1,-3){0.2}}
\multiput(127.33,26.00)(-0.12,-0.35){34}{\line(0,-1){0.35}}
\put(123.33,20.00){\makebox(0,0)[cc]{$i$}}
\put(117.00,54.00){\vector(3,-1){0.2}}
\multiput(65.00,70.00)(0.39,-0.12){134}{\line(1,0){0.39}}
\end{picture}
  \caption{Conditions of theorem \ref{6word}}
  \label{Bilde11}
\end{figure}

\begin{theorem}
\label{6word}
Let $L$ be a language and $M$ be its minimal automaton.
If $M$ contains a fragment of the form shown in Figure \ref{Bilde11}
where $a, b, c, d, e, f, g, h, i\in\Sigma^{*}$ are words and
$q_0$, $q_a$, $q_b$, $q_c$, $q_{ad}$, $q_{ae}$, $q_{bd}$,
$q_{bf}$, $q_{ce}$, $q_{cf}$ are states of $M$ and
\begin{enumerate}
\item
If $M$ reads $x\in \{a, b, c\}$ in the state $q_0$, its state changes to $q_x$.
\item
If $M$ reads $x\in\{a, b, c\}$ in the state $q_x$, its state 
again becomes $q_x$.
\item
If $M$ reads any string consisting of $a$, $b$ and $c$ in a state $q_x$ ($x\in\{a, b, c\}$),
it moves to a state from which it can return to the same state $q_x$ by reading some
(possibly, different) string consisting of $a$, $b$ and $c$.
\item
If $M$ reads $y\in\{d, e, f\}$ in the state $q_x$ ($x\in\{a, b, c\}$), it moves
to the state $q_{xy}$.\footnote{Note: we do not have this constraint 
(and the next two constraints) for pairs $x=a, y=f$, $x=b$, $y=e$ and
$x=c$, $y=d$ for which the state $q_{xy}$ is not defined.}
\item
If $M$ reads $y\in\{a, b, c\}$ in a state $q_{xy}$, its state 
again becomes $q_{xy}$.
\item
If $M$ reads any string consisting of $d$, $e$ and $f$ in the state $q_{xy}$ 
it moves to a state from which it can return to the same state $q_{xy}$ by reading some
(possibly, different) string consisting of $d$, $e$ and $f$.
\item
Reading $g$ in the state $q_{ad}$, $h$ in the state $q_{bf}$ and $i$ in the state $q_{ce}$
leads to accepting states. Reading $h$ in the state $q_{ae}$, $i$ in the state $q_{bd}$, 
$g$ in the state $q_{cf}$ leads to rejecting states.
\end{enumerate}
then $L$ is not recognizable by a QFA.
\end{theorem}

\begin{proof}
For a contradiction, assume that $Q$ is a QFA that recognizes $L$ with a probability 
$1/2+\epsilon$. We construct 6 words such that $Q$ gives a wrong answer on
at least one of them.

Let $\psi$ be a superposition of QFA corresponding to the state $q_0$
(the superposition after reading some word $w$ that leads to $q_0$).
Similarly to the proof of Theorem \ref{C2}, we consider decompositions 
$E_{non}=E_1^{x}\oplus E_2^x$ for all $x\in\{a, b, c\}^*$ and take
$E_1=\cap_{x} E_1^x$, $E_2=E_{non}-E_1$.

Let $\psi=\psi_1+\psi_2$, $\psi_1\in E_1$, $\psi_2\in E_2$.
Similarly to the proof of Theorem \ref{C2}, 
there is a word $a'\in \{a, b, c\}$ with the first letter $a$
such that $\|V'_{a'}(\psi_1)-\psi_1\|\leq \delta$ and
$\|V'_{a'}(\psi_2)\|\leq \delta$ where $\delta=\epsilon/20$.
Also, there are words $b'$ and $c'$ with the first letters $b$ and
$c$ and the same property.

Next, we consider the decompositions $E_{non}=E_1^{x}\oplus E_2^x$ 
for all $x\in\{d, e, f\}^*$ and take $E_3=\cap_{x} E_3^x$, $E_4=E_{non}-E_3$.
Let $\psi_1=\psi_3+\psi_4$, $\psi_3\in E_3$, $\psi_4\in E_4$.
Let $d', e', f'\in\{d, e, f\}^*$ be words with the first letters $d$, $e$ and $f$
such that $\|V'_{d'}(\psi_3)-\psi_3\|\leq\delta$ and 
$\|V'_{d'}(\psi_4)\|\leq \delta$ (and similar inequalities hold 
for $e'$ and $f'$).

Let $p_0$ be the probability of accepting while reading the left endmarker
$\kappa$ and the word $w$ that leads to the superposition $\psi_0$.
Let $p_a$, $p_b$, $p_c$ be the probabilities of accepting while
reading $a'$, $b'$ and $c'$ if the starting superposition is $\psi$.
Let $p_d$, $p_e$ and $p_f$ be the probabilities of accepting while
reading $a'$, $b'$ and $c'$ if the starting superposition is $\psi_1$.
Let $p_g$, $p_h$ and $p_i$ be the probabilities of accepting while
reading $g\$$, $h\$$ and $i\$$ if the starting superposition is $\psi_3$.

\begin{lemma}
Let $p_{a'd'g}$ be the probability of accepting the word $wa'd'g$.
Then,
\[ |p_{a'd'g}-(p_0+p_a+p_d+p_g)| \leq 12\delta.\]
\end{lemma}

\begin{proof}
$p_{a'd'g}$ is the sum of probabilities of accepting while reading $\kappa w$,
accepting while reading $a'$, accepting while reading $d'$ and accepting while reading $g\$$.
The first two probabilities are exactly $p_0$ and $p_a$.

The probability of accepting while reading $d'$ may differ from $p_d$ because the state
of $Q$ after reading $\kappa wa'$ is $V'_{a'}(\psi)$ and the state used to define 
$p_d$ is $\psi_1$. However, these two probabilities differ by at most $4\delta$
because 
\[ \| V'_{a'}(\psi)-\psi_1\|\leq \|V'_{a'}(\psi_1)-\psi_1\|+\|V'_{a'}(\psi_2)\| \leq 2\delta \]
and the probability distributions resulting from observing $V'_{a'}(\psi)$ and $\psi_1$
can differ by at most twice the distance between superpositions (Lemma \ref{LBV}).

Similarly, the distance between the state of $Q$ after reading $\kappa a'd'$ and
$\psi_3$ is at most $4\delta$ and this implies that the probability of accepting
while reading $g\$$ portion of $\kappa wa'd'g\$$ differs from $p_g$ by at most $8\delta$.
Therefore, the difference between $p_{a'd'g}$ and $(p_0+p_a+p_d+p_g)$
is at most $4\delta+8\delta=12\delta$.
\end{proof}

Similar bounds are true for probabilities of accepting $wb'f'h$, $wc'e'i$, $wa'e'h$, 
$wb'd'i$, $wc'f'g$. (We denote these probabilities $p_{b'f'h}$, $p_{c'e'i}$, $p_{a'e'h}$, 
$p_{b'd'i}$, $p_{c'f'g}$.) By putting the bounds for $p_{a'd'g}$, $p_{b'f'h}$, $p_{c'e'i}$
together, we get 
\[ |(p_{a'd'g}+p_{b'f'h}+p_{c'e'i})-(3p_0+p_a+p_b+p_c+p_d+p_e+p_f+p_g+p_h+p_i)| \leq \]
\[ |p_{a'd'g}-(p_0+p_a+p_d+p_g)| + |p_{b'f'h}-(p_0+p_b+p_f+p_h)| + 
|p_{c'e'i}-(p_0+p_c+p_e+p_i)| \leq 36\delta .\]
Putting the bounds for $p_{a'e'h}$, $p_{b'd'i}$, $p_{c'f'g}$ together gives
\[ |(p_{a'e'h}+p_{b'd'i}+p_{c'f'g})-(3p_0+p_a+p_b+p_c+p_d+p_e+p_f+p_g+p_h+p_i)| 
\leq 36 \delta,\]
\[ |(p_{a'd'g}+p_{b'f'h}+p_{c'e'i}) - (p_{a'e'h}+p_{b'd'i}+p_{c'f'g}) | \leq 72\delta .\]
However, each of $p_{a'd'g}$, $p_{b'f'h}$, $p_{c'e'i}$ is the probability of accepting a
word in $L$ and must be at least $1/2+\epsilon$ and each of $p_{a'e'h}$, $p_{b'd'i}$, 
$p_{c'f'g}$ is the probability of accepting a word not in $L$ and must be at most 
$1/2-\epsilon$. Therefore, 
\[ (p_{a'd'g}+p_{b'f'h}+p_{c'e'i}) - (p_{a'e'h}+p_{b'd'i}+p_{c'f'g}) \geq 6\epsilon=120\delta .\]
A contradiction.
\qed
\end{proof}

The existence of the ``forbidden construction'' of Theorem \ref{6word}
does not imply the existence of any of previously shown 
``forbidden constructions''.

\begin{figure}[htb]
  \centering
  \begin{Large}
\special{em:linewidth 0.4pt}
\unitlength 1.00mm
\linethickness{0.4pt}
\begin{picture}(90.00,18.67)
\put(15.00,11.00){\rule{5.00\unitlength}{5.00\unitlength}}
\put(20.00,11.00){\rule{5.00\unitlength}{5.00\unitlength}}
\put(20.00,1.00){\framebox(5.00,5.00)[cc]{}}
\put(25.00,1.00){\framebox(5.00,5.00)[cc]{}}
\put(20.00,6.00){\framebox(5.00,5.00)[cc]{}}
\put(25.00,6.00){\framebox(5.00,5.00)[cc]{}}
\put(25.00,11.00){\framebox(5.00,5.00)[cc]{}}
\put(12.33,14.00){\makebox(0,0)[cc]{$d$}}
\put(12.33,9.00){\makebox(0,0)[cc]{$e$}}
\put(12.33,4.00){\makebox(0,0)[cc]{$f$}}
\put(17.67,18.67){\makebox(0,0)[cc]{$g$}}
\put(22.67,18.67){\makebox(0,0)[cc]{$h$}}
\put(27.67,18.67){\makebox(0,0)[cc]{$i$}}
\put(75.00,1.00){\rule{5.00\unitlength}{5.00\unitlength}}
\put(45.00,1.00){\rule{5.00\unitlength}{5.00\unitlength}}
\put(75.00,6.00){\rule{5.00\unitlength}{5.00\unitlength}}
\put(45.00,6.00){\rule{5.00\unitlength}{5.00\unitlength}}
\put(75.00,11.00){\rule{5.00\unitlength}{5.00\unitlength}}
\put(45.00,11.00){\rule{5.00\unitlength}{5.00\unitlength}}
\put(80.00,11.00){\rule{5.00\unitlength}{5.00\unitlength}}
\put(50.00,11.00){\rule{5.00\unitlength}{5.00\unitlength}}
\put(80.00,1.00){\framebox(5.00,5.00)[cc]{}}
\put(50.00,1.00){\framebox(5.00,5.00)[cc]{}}
\put(85.00,1.00){\framebox(5.00,5.00)[cc]{}}
\put(55.00,1.00){\framebox(5.00,5.00)[cc]{}}
\put(80.00,6.00){\framebox(5.00,5.00)[cc]{}}
\put(85.00,6.00){\framebox(5.00,5.00)[cc]{}}
\put(85.00,11.00){\framebox(5.00,5.00)[cc]{}}
\put(72.33,14.00){\makebox(0,0)[cc]{$d$}}
\put(42.33,14.00){\makebox(0,0)[cc]{$d$}}
\put(72.33,9.00){\makebox(0,0)[cc]{$e$}}
\put(42.33,9.00){\makebox(0,0)[cc]{$e$}}
\put(72.33,4.00){\makebox(0,0)[cc]{$f$}}
\put(42.33,4.00){\makebox(0,0)[cc]{$f$}}
\put(77.67,18.67){\makebox(0,0)[cc]{$g$}}
\put(47.67,18.67){\makebox(0,0)[cc]{$g$}}
\put(82.67,18.67){\makebox(0,0)[cc]{$h$}}
\put(52.67,18.67){\makebox(0,0)[cc]{$h$}}
\put(87.67,18.67){\makebox(0,0)[cc]{$i$}}
\put(57.67,18.67){\makebox(0,0)[cc]{$i$}}
\put(15.00,1.00){\framebox(5.00,5.00)[cc]{}}
\put(15.00,6.00){\framebox(5.00,5.00)[cc]{}}
\put(55.00,11.00){\rule{5.00\unitlength}{5.00\unitlength}}
\put(55.00,6.00){\rule{5.00\unitlength}{5.00\unitlength}}
\put(50.00,6.00){\rule{5.00\unitlength}{5.00\unitlength}}
\put(7.00,9.00){\makebox(0,0)[cc]{$a$:}}
\put(37.00,9.00){\makebox(0,0)[cc]{$b$:}}
\put(67.00,9.00){\makebox(0,0)[cc]{$c$:}}
\end{picture}
  \end{Large}
  \caption{The language $L$}
  \label{Bilde12}
\end{figure}
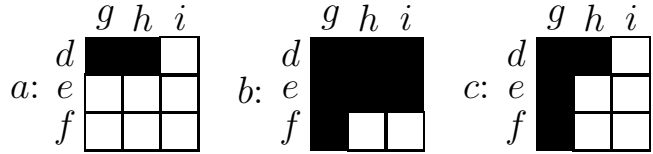

This can be shown as follows.
Consider the alphabet $\Sigma=\{a, b, c, d, e, f, g, h, i\}$ and
languages of the form 
$L_{x, y, z}=x(a|b|c)^* y(d|e|f)^* z$ where
$x\in\{a, b, c\}$, $y\in\{d, e, f\}$, $z\in\{g, h, i\}$.
Our language $L$ will be the union of languages
$L_{x, y, z}$ corresponding to black squares
in Figure \ref{Bilde12}. 

\begin{theorem}
The minimal automaton of $L$ does not contain 
the ``forbidden construction'' of Theorem \ref{C2}.
\end{theorem}

\begin{proof}
The minimal automaton of $L$ has a structure similar to 
Figure \ref{Bilde11}, with some more states.
Similarly to Figure \ref{Bilde11},  
the states of the minimal automaton of $L$
can be partitioned into 4 levels: 
\begin{enumerate}
\item
The starting state (nothing read so far).
\item
The states after reading $a$, $b$ or $c$.
\item
The states after reading $a$, $b$ or $c$
and $d$, $e$ or $f$. 
\item
The states after reading $a$, $b$ or $c$,
$d$, $e$ or $f$ and $g$, $h$ or $i$.
\end{enumerate}

If we look for the ``forbidden construction'' of Theorem \ref{C2},
then $q_2$ and $q_3$ should be in the $2^{\rm nd}$ or $3^{\rm rd}$ 
level. 
($q_2$ or $q_3$ cannot be in the $1^{\rm st}$ level because, after 
reading any letter in the starting state, $M$ leaves it and never  
returns. Also, $q_2$ or $q_3$ cannot be in the $4^{\rm th}$ level
because every state in it is ``all-accepting'' or ``all-rejecting''.)
This leaves us with 3 possible cases.

\begin{enumerate}
\item
$q_2$ and $q_3$ are both states on the $2^{\rm nd}$ level
(after the automaton has read $a$, $b$ or $c$).

The sets of words that are accepted from $q_2$ and $q_3$
correspond to black pieces in $3\times 3$ squares in Figure \ref{Bilde12}. 
For any two of these three squares,
the black pieces in one of them are subset of the black pieces in the other.
(And this means there is no words $z_1, z_2$ such that $z_1$ gets accepted
from $q_2$ but not from $q_3$ and $z_2$ gets accepted from $q_3$ and 
not from $q_2$.) 

\item
$q_2$ and $q_3$ are two of the 9 states on the $3^{\rm rd}$ level
(after reading one of $a$, $b$ or $c$ and one of $d$, $e$ or $f$).

The sets of words that lead to acceptance correspond 
to rows in Figure \ref{Bilde12}. 
One can easily see that any two of them are subsets of one another.

\item
One of $q_2$ and $q_3$ is on the $2^{\rm nd}$ level and the other is on the
$3^{\rm rd}$ level.

W. l. o. g., assume that $q_2$ is on the $2^{\rm nd}$ level and
$q_3$ is on the $3^{\rm rd}$ level.
Then, the word $y$ that leads the automaton $M$ from $q_1$ to $q_3$
must contain one of letters $d$, $e$ and $f$.
However, reading $d$, $e$ or $f$ in the state $q_2$ would lead
$M$ to a state in the $3^{\rm rd}$ level from which it
cannot return to $q_2$ (and, therefore, the condition 6 of Theorem
\ref{C2} is violated).
\end{enumerate}

In all 3 cases, we see that one of conditions of Theorem \ref{C2}
is violated. Therefore, the minimal automaton $M$ does not
contain the ``forbidden construction" of Theorem \ref{C2}.
\qed
\end{proof}

However, one can easily see that the minimal automaton of $L$ contains
the ``forbidden construction'' of Theorem \ref{6word}.
(Just take $q_0$ to be the starting state and make $a$, $b$, $\ldots$, $i$
of Theorem \ref{6word} equal to corresponding letters in the alphabet $\Sigma$.)
This means that the existence of ``forbidden construction'' 
of Theorem \ref{6word} does not imply the existence of the 
``forbidden construction'' of Theorem \ref{C2}.

Theorem \ref{6word} can be generalized to any number of levels
(cycles following one another) and any number of branchings at one
level as long as every arc from one vertex to other is traversed
the same number of times in paths leading to accepting states 
and in paths leading to rejecting states.

A general ``forbidden construction" is as follows.\\

Level 1 of such a construction consists of a state $q_1$ and some words 
$a_{11}$, $a_{12}$, $\ldots$.

Level 2 consists of the states $q_{21}$, $q_{22}$, $\ldots$
where the automaton goes if it reads one of words of Level 1 in
a state in Level 1.
We require that, if the automaton starts in one of states 
of Level 2 and reads any string consisting of 
words of Level 1 it can return to the same
state reading some string consisting of these words.
Level 2 also has some words $a_{21}$, $a_{22}$, $\ldots$.

Level 3 consists of the states $q_{31}$, $q_{32}$, $\ldots$
where the automaton goes if it reads one of words of Level 2 in
a state in Level 2.
We require that, if the automaton starts in one of states of
Level 3 and reads 
any string consisting of 
words of Level 2 it can return to the same
state reading some string consisting of these words.
Again, Level 3 also has some words $a_{31}$, $a_{32}$, $\ldots$.
...

Level $n$ consists of the states $q_{n1}$, $q_{n2}$, $\ldots$
where the automaton goes if it reads one of words of Level $n-1$ in
a state in Level $n-1$.\\


Let us denote all different words in this construction as 
$a_1, a_2, a_3, \ldots , a_m$.

For a word $a_i$ and a level $j$ we construct sets of states $B_{ij}$ 
and $D_{ij}$.
A state $q$ in level $j+1$
belongs to $B_{ij}$ if the word $a_i$ belongs to level $j$
and $M$ moves to $q$ after reading $a_i$ in some state in level $j$.
A state belongs to $D_{ij}$ if this state belongs to the Level $n$ and 
it is reachable from $B_{ij}$.

\begin{theorem}
Assume that the minimal automaton $M$ of a language $L$ contains 
the ``forbidden construction" of the general form described above
and, in this construction, for each $D_{ij}$
the number of accepting states is equal to the number of rejecting states.  
Then, $L$ cannot be recognized by a 1-way QFA.
\end{theorem}

Theorems \ref{C2} and \ref{6word} are special cases of
this theorem (with 3 and 4 levels, respectively).

\begin{figure}
  \setlength{\abovecaptionskip}{0pt}
  \begin{minipage}{0.5\linewidth}
    \centering
\special{em:linewidth 0.4pt}
\unitlength 1.00mm
\linethickness{0.4pt}
\begin{picture}(50.00,45.00)
\put(50.00,5.00){\vector(1,0){0.2}}
\emline{10.00}{5.00}{1}{50.00}{5.00}{2}
\put(10.00,45.00){\vector(0,1){0.2}}
\emline{10.00}{5.00}{3}{10.00}{45.00}{4}
\emline{16.00}{5.00}{5}{16.00}{17.00}{6}
\emline{16.00}{17.00}{7}{10.00}{17.00}{8}
\emline{34.00}{5.00}{9}{34.00}{17.00}{10}
\emline{40.00}{17.00}{11}{40.00}{5.00}{12}
\emline{10.00}{23.00}{13}{16.00}{23.00}{14}
\emline{16.00}{35.00}{15}{10.00}{35.00}{16}
\emline{40.00}{23.00}{17}{40.00}{35.00}{18}
\emline{40.00}{17.00}{19}{40.00}{23.00}{20}
\emline{34.00}{17.00}{21}{34.00}{23.00}{22}
\emline{34.00}{23.00}{23}{16.00}{23.00}{24}
\emline{34.00}{35.00}{25}{40.00}{35.00}{26}
\emline{34.00}{35.00}{27}{16.00}{35.00}{28}
\begin{large}
\put(13.00,11.00){\makebox(0,0)[cc]{$S_1$}}
\put(37.00,29.00){\makebox(0,0)[cc]{$S_2$}}
\end{large}
\emline{-1.33}{31.00}{29}{45.00}{-3.67}{30}
\put(7.00,2.00){\makebox(0,0)[cc]{$0$}}
\put(16.00,2.00){\makebox(0,0)[cc]{$1\!-\!p_1$}}
\put(34.00,2.00){\makebox(0,0)[cc]{$p_1$}}
\put(40.00,2.00){\makebox(0,0)[cc]{$1$}}
\put(49.00,2.00){\makebox(0,0)[cc]{$x$}}
\put(7.00,17.00){\makebox(0,0)[rc]{$1\!-\!p_2\!\!$}}
\put(7.00,23.00){\makebox(0,0)[cc]{$p_2$}}
\put(7.00,35.33){\makebox(0,0)[cc]{$1$}}
\put(7.00,44.00){\makebox(0,0)[cc]{$y$}}
\end{picture}

    \setlength{\abovecaptionskip}{0pt}
    \setlength{\belowcaptionskip}{30pt}
    \caption{}
    \label{Bilde5}
  \end{minipage}%
  \begin{minipage}{0.5\linewidth}
    \centering
\special{em:linewidth 0.4pt}
\unitlength 1.00mm
\linethickness{0.4pt}
\begin{picture}(50.00,45.00)
\put(50.00,5.00){\vector(1,0){0.2}}
\emline{10.00}{5.00}{1}{50.00}{5.00}{2}
\put(10.00,45.00){\vector(0,1){0.2}}
\emline{10.00}{5.00}{3}{10.00}{45.00}{4}
\emline{40.00}{17.00}{5}{40.00}{5.00}{6}
\emline{16.00}{35.00}{7}{10.00}{35.00}{8}
\emline{40.00}{23.00}{9}{40.00}{35.00}{10}
\emline{40.00}{17.00}{11}{40.00}{23.00}{12}
\emline{34.00}{35.00}{13}{40.00}{35.00}{14}
\emline{34.00}{35.00}{15}{16.00}{35.00}{16}
\begin{large}
\put(16.00,11.00){\makebox(0,0)[cc]{$S_1$}}
\put(33.00,28.00){\makebox(0,0)[cc]{$S_2$}}
\end{large}
\put(7.00,2.00){\makebox(0,0)[cc]{$0$}}
\put(22.00,2.00){\makebox(0,0)[cc]{$1\!-\!p_1$}}
\put(28.00,2.00){\makebox(0,0)[cc]{$p_1$}}
\put(40.00,2.00){\makebox(0,0)[cc]{$1$}}
\put(49.00,2.00){\makebox(0,0)[cc]{$x$}}
\put(7.00,17.00){\makebox(0,0)[rc]{$1\!-\!p_2\!\!$}}
\put(7.00,23.00){\makebox(0,0)[cc]{$p_2$}}
\put(7.00,35.33){\makebox(0,0)[cc]{$1$}}
\put(7.00,44.00){\makebox(0,0)[cc]{$y$}}
\emline{22.00}{5.00}{17}{22.00}{17.00}{18}
\emline{22.00}{17.00}{19}{10.00}{17.00}{20}
\emline{28.00}{5.00}{21}{28.00}{23.00}{22}
\emline{28.00}{23.00}{23}{10.00}{23.00}{24}
\emline{5.00}{30.00}{25}{38.00}{-3.00}{26}
\end{picture}

    \setlength{\abovecaptionskip}{0pt}
    \setlength{\belowcaptionskip}{30pt}
    \caption{}
    \label{Bilde6}
  \end{minipage}\\
  \begin{minipage}{0.5\linewidth}
    \centering
\special{em:linewidth 0.4pt}
\unitlength 1.00mm
\linethickness{0.4pt}
\begin{picture}(50.00,45.00)
\put(50.00,5.00){\vector(1,0){0.2}}
\emline{10.00}{5.00}{1}{50.00}{5.00}{2}
\put(10.00,45.00){\vector(0,1){0.2}}
\emline{10.00}{5.00}{3}{10.00}{45.00}{4}
\emline{40.00}{17.00}{5}{40.00}{5.00}{6}
\emline{16.00}{35.00}{7}{10.00}{35.00}{8}
\emline{40.00}{23.00}{9}{40.00}{35.00}{10}
\emline{40.00}{17.00}{11}{40.00}{23.00}{12}
\emline{34.00}{35.00}{13}{40.00}{35.00}{14}
\emline{34.00}{35.00}{15}{16.00}{35.00}{16}
\begin{large}
\put(15.00,10.00){\makebox(0,0)[cc]{$S_1$}}
\put(34.00,29.00){\makebox(0,0)[cc]{$S_2$}}
\end{large}
\put(7.00,2.00){\makebox(0,0)[cc]{$0$}}
\put(7.00,35.33){\makebox(0,0)[cc]{$1$}}
\put(7.00,44.00){\makebox(0,0)[cc]{$y$}}
\emline{20.00}{5.00}{17}{20.00}{15.00}{18}
\emline{20.00}{15.00}{19}{10.00}{15.00}{20}
\emline{30.00}{5.00}{21}{30.00}{25.00}{22}
\emline{30.00}{25.00}{23}{10.00}{25.00}{24}
\emline{7.00}{28.00}{25}{35.00}{0.00}{26}
\put(20.33,2.00){\makebox(0,0)[cc]{$\frac{1}{3}$}}
\put(30.33,2.00){\makebox(0,0)[cc]{$\frac{2}{3}$}}
\put(40.33,2.00){\makebox(0,0)[cc]{$1$}}
\put(7.33,14.00){\makebox(0,0)[cc]{$\frac{1}{3}$}}
\put(7.33,24.00){\makebox(0,0)[cc]{$\frac{2}{3}$}}
\end{picture}

    \setlength{\abovecaptionskip}{0pt}
    \setlength{\belowcaptionskip}{30pt}
    \caption{}
    \label{Bilde9}
  \end{minipage}%
  \begin{minipage}{0.5\linewidth}
    \centering
\special{em:linewidth 0.4pt}
\unitlength 1.00mm
\linethickness{0.4pt}
\begin{picture}(50.00,45.00)
\put(50.00,5.00){\vector(1,0){0.2}}
\emline{10.00}{5.00}{1}{50.00}{5.00}{2}
\put(10.00,45.00){\vector(0,1){0.2}}
\emline{10.00}{5.00}{3}{10.00}{45.00}{4}
\begin{large}
\put(20.00,15.00){\makebox(0,0)[cc]{$S_1$}}
\put(30.00,25.00){\makebox(0,0)[cc]{$S_2$}}
\end{large}
\put(7.00,2.00){\makebox(0,0)[cc]{$0$}}
\put(22.00,2.00){\makebox(0,0)[cc]{$1\!-\!p_1$}}
\put(28.00,2.00){\makebox(0,0)[cc]{$p_1$}}
\put(40.00,2.00){\makebox(0,0)[cc]{$1$}}
\put(49.00,2.00){\makebox(0,0)[cc]{$x$}}
\put(7.00,17.00){\makebox(0,0)[rc]{$1\!-\!p_2\!\!$}}
\put(7.00,23.00){\makebox(0,0)[cc]{$p_2$}}
\put(7.00,35.33){\makebox(0,0)[cc]{$1$}}
\put(7.00,44.00){\makebox(0,0)[cc]{$y$}}
\emline{22.00}{17.00}{5}{22.00}{13.00}{6}
\emline{22.00}{13.00}{7}{18.00}{13.00}{8}
\emline{18.00}{13.00}{9}{18.00}{17.00}{10}
\emline{18.00}{17.00}{11}{22.00}{17.00}{12}
\emline{28.00}{13.00}{13}{32.00}{13.00}{14}
\emline{32.00}{13.00}{15}{32.00}{27.00}{16}
\emline{32.00}{27.00}{17}{18.00}{27.00}{18}
\emline{18.00}{27.00}{19}{18.00}{23.00}{20}
\emline{18.00}{23.00}{21}{28.00}{23.00}{22}
\emline{28.00}{23.00}{23}{28.00}{13.00}{24}
\emline{8.00}{32.00}{25}{41.00}{0.00}{26}
\end{picture}

    \setlength{\abovecaptionskip}{0pt}
    \setlength{\belowcaptionskip}{30pt}
    \caption{}
    \label{Bilde7}
  \end{minipage}\\
  \centering
\special{em:linewidth 0.4pt}
\unitlength 1.00mm
\linethickness{0.4pt}
\begin{picture}(50.00,45.00)
\put(50.00,5.00){\vector(1,0){0.2}}
\emline{10.00}{5.00}{1}{50.00}{5.00}{2}
\put(10.00,45.00){\vector(0,1){0.2}}
\emline{10.00}{5.00}{3}{10.00}{45.00}{4}
\begin{large}
\put(20.00,15.00){\makebox(0,0)[cc]{$S_1$}}
\put(30.00,25.00){\makebox(0,0)[cc]{$S_2$}}
\end{large}
\put(7.00,2.00){\makebox(0,0)[cc]{$0$}}
\put(22.00,2.00){\makebox(0,0)[cc]{$1\!-\!p_1$}}
\put(28.00,2.00){\makebox(0,0)[cc]{$p_1$}}
\put(40.00,2.00){\makebox(0,0)[cc]{$1$}}
\put(49.00,2.00){\makebox(0,0)[cc]{$x$}}
\put(7.00,17.00){\makebox(0,0)[rc]{$1\!-\!p_2\!\!$}}
\put(7.00,23.00){\makebox(0,0)[cc]{$p_2$}}
\put(7.00,35.33){\makebox(0,0)[cc]{$1$}}
\put(7.00,44.00){\makebox(0,0)[cc]{$y$}}
\emline{9.00}{32.67}{5}{42.00}{0.67}{6}
\emline{40.00}{4.00}{7}{40.00}{6.00}{8}
\emline{9.00}{35.00}{9}{11.00}{35.00}{10}
\emline{28.00}{4.00}{11}{28.00}{6.00}{12}
\emline{22.00}{4.00}{13}{22.00}{6.00}{14}
\emline{9.00}{17.00}{15}{11.00}{17.00}{16}
\emline{9.00}{23.00}{17}{11.00}{23.00}{18}
\emline{28.00}{17.00}{19}{28.00}{23.00}{20}
\emline{28.00}{23.00}{21}{22.00}{23.00}{22}
\put(22.00,17.00){\circle*{0.67}}
\end{picture}

  \setlength{\abovecaptionskip}{0pt}
  \setlength{\belowcaptionskip}{30pt}
  \caption{}
  \label{Bilde8}
\end{figure}

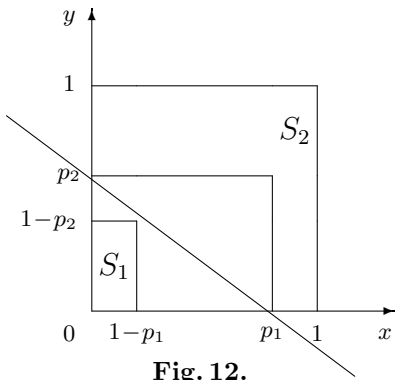
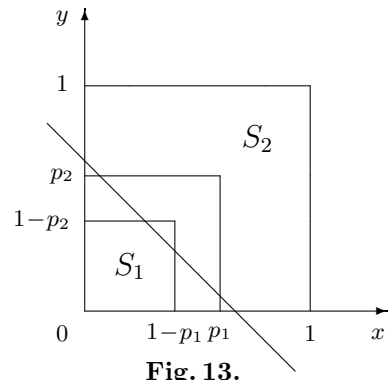
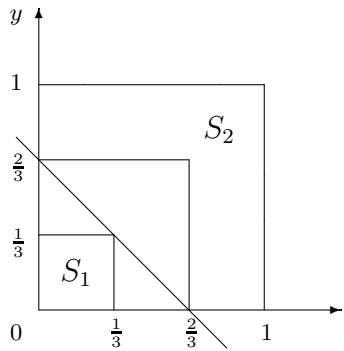
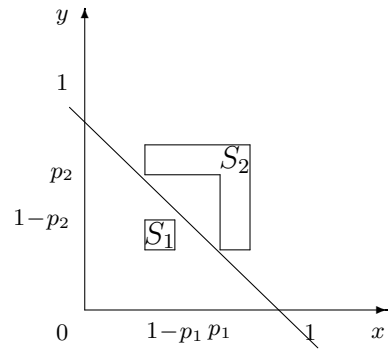
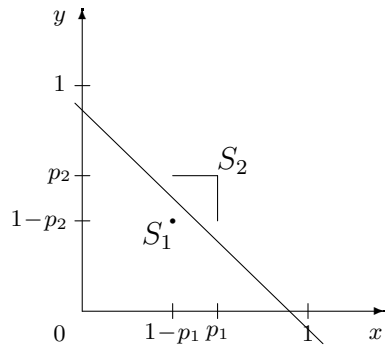
\end{document}